\documentclass[%
 reprint,
superscriptaddress,
 amsmath,amssymb,
aps,
longbibliography,
pra,
]{revtex4-1}

\usepackage{graphicx}
\usepackage{dcolumn}
\usepackage{bm}

\usepackage{times}
\usepackage{subcaption,bbm}
\usepackage{physics}
\allowdisplaybreaks
\usepackage[dvipsnames]{xcolor}

\newcommand{\ii}{\mathrm{i}}
\newcommand{\ee}{\mathrm{e}}

\newcommand{\dirac}{\mathrm{D}}
\newcommand{\sch}{\mathrm{S}}
\newcommand{\lk}{\ket{\mathrm{L}}}

\newcommand{\rk}{\ket{\mathrm{R}}}

\newcommand{\dk}{\ket{\mathrm{D}}}

\newcommand{\uk}{\ket{\mathrm{U}}}

\begin{document}

\title{Multi-Dimensional Quantum Walks:\\ a Playground of Dirac and Schr\"{o}dinger Particles}
\author{Manami Yamagishi}
\email{manami@iis.u-tokyo.ac.jp}
\affiliation{Department of Physics, the University of Tokyo,
5-1-5 Kashiwanoha, Kashiwa, Chiba 277-8574, Japan}
\author{Naomichi Hatano}
\email{hatano@iis.u-tokyo.ac.jp}
\author{Ken-Ichiro Imura}
\email{imura@iis.u-tokyo.ac.jp}
\affiliation{Institute of Industrial Science, the University of Tokyo,
5-1-5 Kashiwanoha, Kashiwa, Chiba 277-8574, Japan}
\author{Hideaki Obuse}
\email{hideaki.obuse@eng.hokudai.ac.jp}
\affiliation{Department of Applied Physics, Hokkaido University,
Kita 13, Nishi 8, Kita-Ku, Sapporo, Hokkaido 060-8628, Japan}
\affiliation{Institute of Industrial Science, the University of Tokyo,
5-1-5 Kashiwanoha, Kashiwa, Chiba 277-8574, Japan}

\date{\today}

\begin{abstract}
We propose a new multi-dimensional discrete-time quantum walk (DTQW), whose continuum limit is an extended multi-dimensional Dirac equation, 
which can be further mapped to the Schr\"{o}dinger equation.
We show in two ways that our DTQW is an excellent measure to investigate the two-dimensional (2D) extended Dirac Hamiltonian and higher-order topological materials.
First, we show that the dynamics of our DTQW resembles that of a 2D Schr\"{o}dinger harmonic oscillator.
Second, we find in our DTQW topological features of the extended Dirac system.
By manipulating the coin operators, we can generate not only standard edge states but also corner states. 
\end{abstract}
\maketitle


\section{\label{sec1}Introduction}
%
The quantum walk is a quantum analogue of random walk.
Instead of stochastic fluctuations of a classical random walker, a quantum walker moves under interference of quantum fluctuations at each site, which deterministically governs the walker's dynamics.
Quantum walk was originally introduced by Aharonov \textit{et al.}~\cite{Aharonov93}, who first referred to it as ``quantum random walk.''
Meyer~\cite{Meyer96} built a systematic model and found a correspondence to Feynman's path integral~\cite{Feynman65} of the Dirac equation.
Started by Farhi and Gutmann~\cite{Farhi98}, quantum walks have been well studied in the context of quantum information~\cite{Ambainis12, Asaka21}.
To this day, studies of quantum walks have become even more interdisciplinary and extended over a variety of research fields, such as biophysics~\cite{Engel07, Dudhe22} and condensed-matter physics~\cite{Oka05}, particularly topological materials~\cite{Kitagawa10, Kitagawa12, Asboth13}.

There are two types of time evolution: continuum-time quantum walks and discrete-time quantum walks (DTQW).
We focus on the latter, in which the space and time are both discrete.
Strauch~\cite{Strauch06} showed that the continuum limit of the unitary time evolution of one-dimensional (1D) DTQW gives that of a Dirac particle.
This correspondence of DTQW has enabled us to understand physical meaning of quantum walks better.
Since squaring the Dirac Hamiltonian with a linear potential produces the Schr\"{o}dinger Hamiltonian with a harmonic potential, we can make further correspondence between a quantum walker and a Schr\"{o}dinger particle in 1D.
However, such investigation has been limited to 1D systems.
In two-dimensional (2D) systems, some quantum walks give a Dirac Hamiltonian in its continuum limit~\cite{Franco11, Bru16, Arrighi18}, but there are usually only two internal states and hence one cannot obtain the 2D Schr\"{o}dinger Hamiltonian by squaring it.
\begin{figure}[t]
  \centering
	\includegraphics[width=\columnwidth]{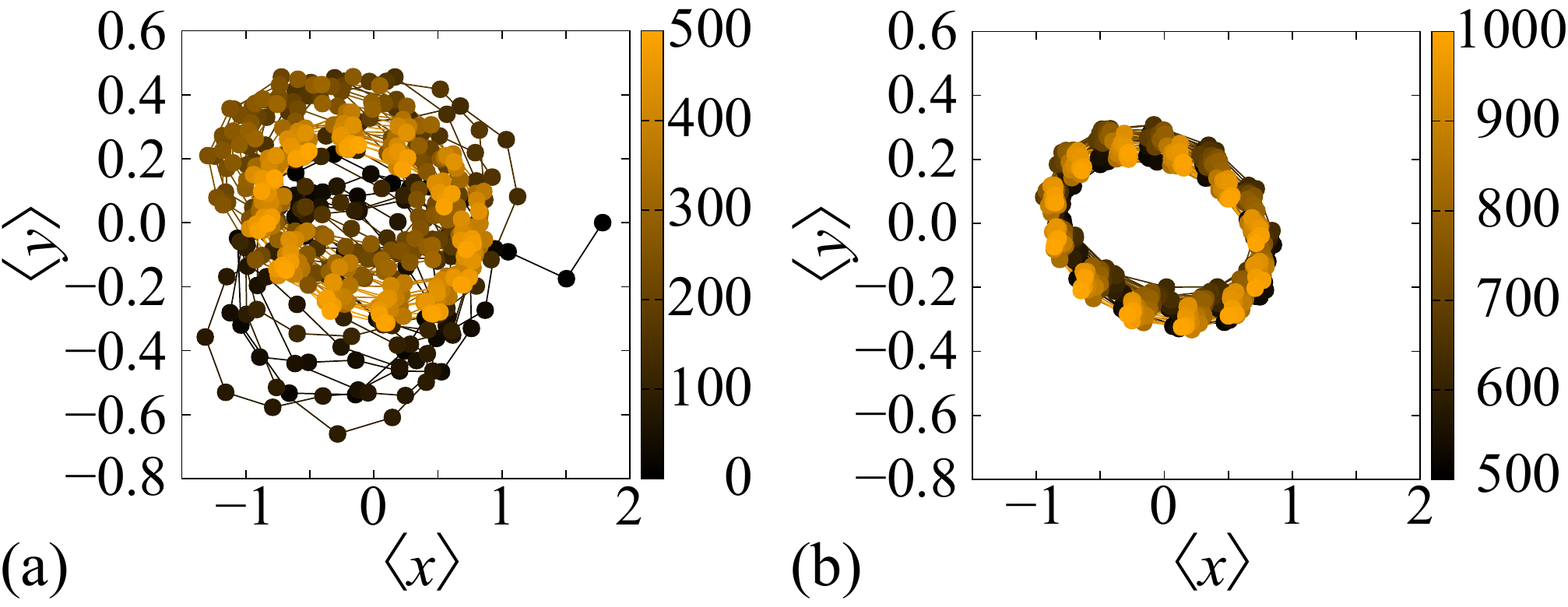}
  \caption{(Color online). The expectation values $\expval{x}$ and $\expval{y}$ of the position of 2D DTQW for (a) $0\le T\le500$ and for (b) $500\le T\le1000$. Black circles indicate the values at the beginning of time evolution; they turn into orange as time goes on. We set $\hbar=a=\Delta t=1$.}
\label{orbit_1000}
\vspace{2\baselineskip}
  \centering
	\includegraphics[width=\columnwidth]{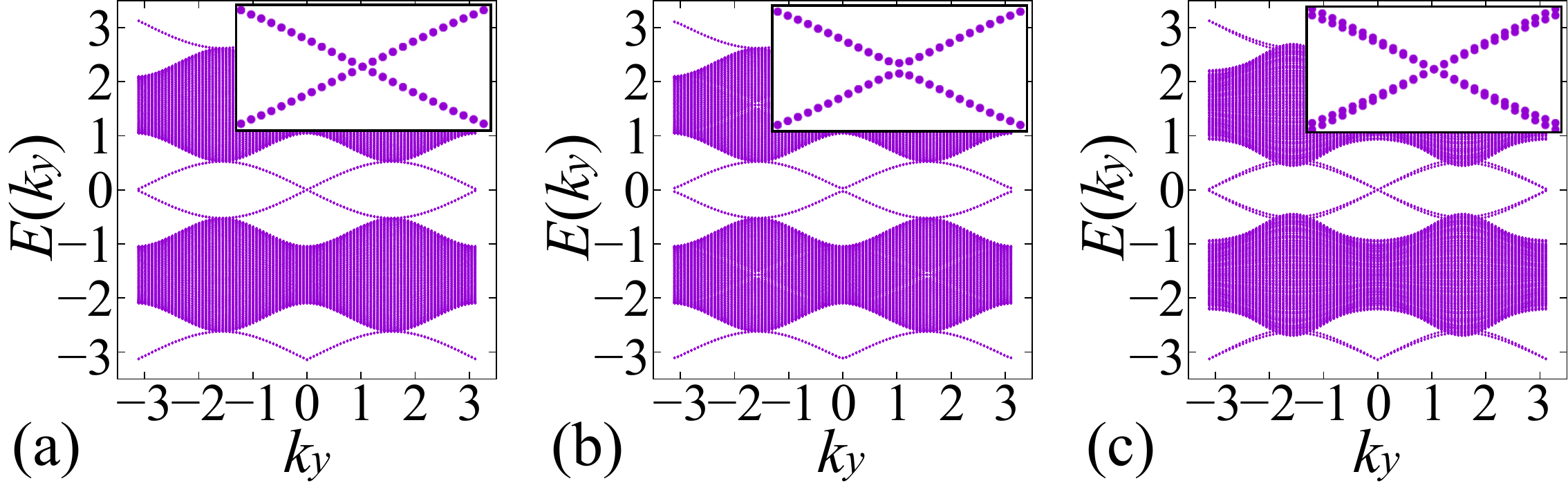}
  \caption{Dispersion relation of the quasi-energy spectra $E(k_y)$ with $\theta_1=-\theta_2=\pi/3$.
 (a) $\theta_y=0$ and (b) $\theta_y=\pi/50$ without randomness, and (c) $\theta_y=0$ with randomness $\Delta{\theta_x}_0(x)\in[-0.25, 0.25]$. The central part of the dispersion is enlarged in the upper right corner in each panel. We set $\hbar=a=\Delta t=1$.}
  \label{bands}
\end{figure}

In this paper, we show in two ways that our DTQW is an excellent measure to investigate an extended 2D Dirac Hamiltonian with additional internal degrees of freedom.
We analyze its dynamics and topological properties, especially higher-order topological insulators~\cite{Murani17,Imhof18,Peterson18}.
We start with proposing a 2D extended Dirac Hamiltonian
\begin{align}\label{H_2D}
&H_{\dirac}^{(2)}
:=H_{\dirac_x}\otimes\tau^0+\sigma^x\otimes H_{\dirac_y} \notag \\
&:=\qty(\epsilon\sigma^zp_x+m_x(x)\sigma^y)\otimes\tau^0
+\sigma^x\otimes\qty(\epsilon\tau^zp_y+m_y(y)\tau^y).
\end{align}
that can be mapped to a 2D DTQW as well as to a Schr\"{o}dinger Hamiltonian as we will show below.
Our key trick is to introduce $\sigma^x$ in the second term so that upon squaring the Hamiltonian~\eqref{H_2D} all crossing terms may vanish and the result can be the standard 2D Schr\"{o}dinger Hamiltonian.
With the additional internal degrees of freedom due to the introduction of $\sigma^x$, we refer to the Dirac Hamiltonian~\eqref{H_2D} as the extended Dirac Hamiltonian.
The same applies to higher-dimensional cases; see Eq.~\eqref{eq31} below for the three-dimensional (3D) case with the eight-dimensional internal degree of freedom.

Remember that the original 3D Dirac equation is written with a four-dimensional spinor degree of freedom, which Dirac assigned to a particle and an anti-particle each with a spin 1/2 degree of freedom. 
This is partly because the four-dimensional degree of freedom is the minimal representation of gamma matrices that satisfies the anticommutation relation necessary in deriving the Dirac equation from the Klein-Gordon equation; see \textit{e.g.}\ Ref.~\cite{Ryder96}. 
However, it does not mean that we cannot go beyond Dirac’s minimal representation. 
As far as the dimensionality of the spinor degree of freedom is a multiple of four in the spatial three-dimensional case as in Eq.~\eqref{eq31} below, it is possible to construct Dirac-type equations with additional internal degrees of freedom, which is what we have done here. 
Thus we refer to our model as the extended Dirac Hamiltonian.

As another point, we might have had to refer to the square of the Dirac Hamiltonian as the Klein-Gordon equation, but we will below refer to it as the Schr\"{o}dinger Hamiltonian because we will demonstrate the behavior of a harmonic oscillator under the linear spacial dependence of the mass terms $m_x(x)$ and $m_y(y)$. 
Indeed, the existence of the two mass terms in Eq.~\eqref{H_2D} is also a new feature of our extended Dirac Hamiltonian.

The introduction of $\sigma^x$ in Eq.~\eqref{H_2D} also enables corner states, the second-order topological states, to emerge.
The higher-order topological states has come under an intensive investigation in these days (see \textit{e.g.} Refs.~\cite{Murani17,Imhof18,Peterson18,BBH,Hayashi18,Schindler,trif,ZF,BBH2}).
A systematic construction of Hamiltonians that harbors higher-order topological states has been developed recently by Hayashi~\cite{Hayashi18}.
The extended Dirac Hamiltonian proposed in the present paper turns out to follow the construction of higher-order topological states, and hence the present two-dimensional DTQW explicitly exhibits corner states.
In other words, the present DTQW models simulate quantum dynamics of higher-order topological insulators.

We numerically find that our 2D quantum walker behaves like a 2D harmonic oscillator as shown in Fig.~\ref{orbit_1000}, to which we will get back below.
We also reveal nontrivial topological properties of our DTQW using the implication of the Dirac Hamiltonian~\cite{Jackiw76}.
We also numerically find two different types of topological bound states, namely edge states of the topology of type $2\mathbb{Z}$ (which are robust against randomness in Fig.~\ref{bands}) and corner states, by manipulating the coin operators of our DTQW.
(See below for the definitions of the notations in Figs.~\ref{orbit_1000} and~\ref{bands}.)

\subsection{\label{sec1A}Review of One-Dimensional Case}
Let us first review the continuum limit of the 1D DTQW.
We define the time evolution of the standard 1D quantum walk $\ket{\psi(T)}={U^{(1)}}^T\ket{\psi(0)}$ for $T\in\mathbbm{Z}$ in terms of the following coin and shift operators:
\begin{align}
C&:=\sum_{x\in a\mathbbm{Z}}\dyad{x}\otimes\ee^{-\ii\theta_x(x)\sigma^y}, \label{eq10} \\
S&:=\sum_{x\in a\mathbbm{Z}}(\dyad{x-a}{x}\otimes\dyad{\mathrm{L}}+\dyad{x+a}{x}\otimes\dyad{\mathrm{R}})
\label{eq11}
\end{align}
with $U^{(1)}:=SC$.
Here, $\theta_x(x)$ is a coefficient set to a linear function of $x$ below, $a$ is the lattice constant and $\{\sigma^x, \sigma^y, \sigma^z\}$ are the Pauli matrices in the space spanned by the leftward state, $\lk=(1, 0)^{\top}$, and the rightward state, $\rk=(0, 1)^{\top}$.
We set $\hbar$ to unity throughout the paper.

Let us express the shift operator~\eqref{eq11} in the form 
\begin{align}
S=\exp\qty(-a\sigma^z\dv{x})=\exp(-\ii a\sigma^z p_x)
\end{align}
with $p_x=-i\dv*{x}$.
Scaling the parameters $a$ and $\theta_x(x)$  as in
$\epsilon:=a/\Delta t$ and $m_x(x):=\theta_x(x)/\Delta t$
with $t:=T\Delta t$ and taking the limit $\Delta t\to 0$ with $T\to\infty$ under a fixed value of $t$,  
we find the continuum limit of the time-evolution operator in the form of the Trotter formula~\cite{Strauch06}
\begin{align}\label{eq18}
\lim_{\Delta t\to 0}{U^{(1)}}^T%
=\lim_{\Delta t\rightarrow0}\qty[\ee^{-\ii \epsilon\Delta t\sigma^zp_x}\ee^{-\ii m_x(x)\Delta t\sigma^y}]^T%
=\ee^{-\ii H_{\dirac}^{(1)}t},
\end{align}
where
\begin{align}\label{H_1D}
H_{\dirac}^{(1)}=\epsilon\sigma^zp_x+m_x(x)\sigma^y
\end{align}
represents the Hamiltonian of a Dirac particle with mass $m_x(x)$ in 1D.
We can analyze its dynamics approximately by squaring it:
\begin{align}\label{eq13}
{\mqty(H_{\dirac}^{(1)})}^2=(\epsilon^2{p_x}^2+m_x(x)^2)\sigma^0-\ii\epsilon\sigma^x[p_x, m_x(x)]=:H_{\sch}^{(1)},
\end{align}
where $\sigma^0$ denotes the $2\times2$ identity matrix for the space spanned by $\lk$ and $\rk$.
Let us assume that $m_x(x)=\theta_x(x)/\Delta t$ is linear in $x$ as in
$\theta_x(x)=bx$ and $m_x(x)=\beta x$
with $\beta=b/\Delta t$.
This reduces the last term of $H_{\sch}^{(1)}$ to $-\epsilon\beta\sigma^x$.
A unitary transformation $V=\exp(\ii\sigma^y\pi/4)$ turns the last term further to $\epsilon\beta\sigma^z$, diagonalizing the Hamiltonian $H_{\sch}^{(1)}$ to the two blocks of 
\begin{align}\label{sch1}
\tilde{H}_{\sch_{\pm}}^{(1)}:=\epsilon^2{p_x}^2+\beta^2x^2\pm\epsilon\beta,
\end{align}
each of which is the Schr\"{o}dinger Hamiltonian in a 1D harmonic potential with a constant term under the following identification:
\begin{align}\label{eq170}
\epsilon^2\leftrightarrow\frac{1}{2m_{\sch}},\quad
\beta^2\leftrightarrow\frac{m_{\sch}\omega^2}{2},\quad
\epsilon\beta\leftrightarrow\frac{\omega}{2}.
\end{align}

The preceding argument shows that the Dirac and Schr\"{o}dinger Hamiltonians share the same eigenvectors.
Indeed, the time evolution of $H_{\dirac}^{(1)}$ is approximately given by $\tilde{H}_{\sch}^{(1)}$.
We can numerically confirm that the Dirac Hamiltonian makes a wave packet oscillate around $x=0$ approximately like a harmonic oscillator.

\section{\label{sec2}Two-Dimensional Model}
Our first point of the paper is to extend the argument to higher dimensions.
There have been two major kinds of 2D DTQW: the Grover walk~\cite{Shenvi03} and an alternative quantum walk\cite{Franco11}.
However, we cannot map either of them to the Schr\"{o}dinger equation.
Instead of these two DTQWs, we here introduce a new DTQW whose continuum limit yields the extended Dirac Hamiltonian~\eqref{H_2D}.
Let $\lk$, $\rk$, $\dk$ and $\uk$ denote the basis vectors for the leftward, rightward, downward, and upward states, respectively.
In Eq.~\eqref{H_2D}, $\{\sigma^x, \sigma^y, \sigma^z\}$ are the the Pauli matrices for the space spanned by $\ket{\mathrm{L}}$ and $\ket{\mathrm{R}}$, while $\{\tau^x, \tau^y, \tau^z\}$ and $\tau^0$ are the Pauli matrices and the identity matrix for the space spanned by $\ket{\mathrm{D}}$ and $\ket{\mathrm{U}}$.
We let $m_x(x)$ and $m_y(y)$ denote the mass terms.
The momenta $p_x$ and $p_y$ can be rewritten in the forms of $-\ii\partial/\partial x$ and $-\ii\partial/\partial y$, respectively.

We can easily confirm that the extended Dirac Hamiltonian~\eqref{H_2D} is by squaring it mapped to the Schr\"{o}dinger Hamiltonian
\begin{align}\label{eq28}
H_{\sch}^{(2)}%
:=H_{\sch_x}\otimes\tau^0+\sigma^0\otimes H_{\sch_y},%
\end{align}
where 
\begin{align}\label{eq29}
H_{\sch_x}&:=(\epsilon^2{p_x}^2+m_x(x)^2)\sigma_0-\ii\epsilon\sigma^x\comm{p_x}{m_x(x)},\\
\label{eq29-1}
H_{\sch_y}&:=(\epsilon^2{p_y}^2+m_y(y)^2)\tau^0-\ii\epsilon\tau^x\comm{p_y}{m_y(y)}.
\end{align}
Assumptions
\begin{align}\label{linearpot0}
m_x(x)=\beta x\quad\mbox{and}\quad m_y(y)=\beta y
\end{align}
reduce Eq.~\eqref{eq28} to
\begin{align}
H_{\sch}^{(2)}&:=(\epsilon^2{p_x}^2+\beta^2x^2-\epsilon\beta\sigma^x)\otimes\tau^0 \notag \\
&+\sigma^0\otimes(\epsilon^2{p_y}^2+\beta^2y^2-\epsilon\beta\tau^x),
\end{align}
which represents a 2D harmonic oscillator under the identification \eqref{eq170}.

We can further extend the argument into the 3D model with the extended Dirac Hamiltonian
\begin{align}\label{eq31}
H_{\dirac}^{(3)}
&:=H_{\dirac_x}\otimes\tau^0\otimes v^0+\sigma^x\otimes H_{\dirac_y}\otimes v^0+\sigma^x\otimes\tau^x\otimes H_{\dirac_z}
\end{align}
although it may not be a standard 3D Dirac Hamiltonian because we have now $2\times2\times2$ degrees of freedom at each site.
In Eq.~\eqref{eq31}, $v^0$ is the identity matrix for the space spanned by the backward state $\ket{\mathrm{B}}$ and the forward state $\ket{\mathrm{F}}$ of the additional inner degree of freedom, while $\{v^x, v^y, v^z\}$ are the Pauli matrices for the same space.
Extension to even higher dimensions should be obvious.

\subsection{\label{sec2A}Two-Dimensional Oscillator}
We next construct our DTQW model from the extended Dirac Hamiltonian~\eqref{H_2D}.
The Hilbert space for the inner degrees of freedom at each site is now spanned by 
\begin{align}
(\ket{\mathrm{L}}+\ket{\mathrm{R}})\oplus(\ket{\mathrm{D}}+\ket{\mathrm{U}})=\ket{\mathrm{LD}}+\ket{\mathrm{RD}}+\ket{\mathrm{LU}}+\ket{\mathrm{RU}}.
\end{align}
We hereafter fix the ordering of the basis vectors in this way.
After conducting the Trotter decomposition on $\exp(-\ii H_{\dirac}^{(2)}t)$, we obtain the time-evolution operator $U^{(2)}$ in the form of $U^{(2)}=S_yC_yS_xC_x$ with
\begin{align}\label{eq22}
\mqty{
C_x:=\ee^{-\ii \theta_x(x)(\sigma^y\otimes\tau^0)},\quad
S_x:=\ee^{-a(\sigma^z\otimes\tau^0)\partial_x},\\
C_y:=\ee^{-\ii \theta_y(y)(\sigma^x\otimes\tau^y)},\quad
S_y:=\ee^{-a(\sigma^x\otimes\tau^z)\partial_y}.
}
\end{align}

Let us here assume that $\theta_x(x)$ and $\theta_y(y)$ are linear in $x$ and $y$, respectively, as in 
\begin{align}\label{linearpot1}
\theta_x(x)=bx\quad\mbox{and}\quad\theta_y(y)=by,
\end{align}
which are related to Eq.~\eqref{linearpot0} as in $\theta_x(x)=m_x(x)\Delta t$ and $\theta_y(y)=m_y(y)\Delta t$.
We can regard this as effective linear potentials for the corresponding Dirac particle.
The operators $C_x$
and $S_x$ in the $x$ direction are given by straightforwardly extending the corresponding operators~\eqref{eq10} and~\eqref{eq11} for the 1D DTQW, respectively.
On the other hand, the operators $C_y$ and $S_y$ read
\begin{align}\label{eq23}
C_y=\mqty(
+c & & & -s \\
 & +c & -s & \\
 & +s & +c & \\
+s & & & +c
),\quad
S_y=\mqty(
P & Q & & \\
Q & P & & \\
 & & P & -Q \\
 & & -Q & P
),
\end{align}
where
\begin{align}\label{eq24}
\mqty{
&c:=\cos{(by)},\quad s:=\sin{(by)}, \\[2pt]
&P:=\frac{1}{2}(\dyad{y-a}{y}+\dyad{y+a}{y}), \\[2pt]
&Q:=\frac{1}{2}(\dyad{y-a}{y}-\dyad{y+a}{y}).
}
\end{align}
These coin and shift operators in Eq.~\eqref{eq23} look differently from the Grover walk~\cite{Shenvi03} and the alternative quantum walk~\cite{Franco11} because of the $\sigma^x$ term in the extended Dirac Hamiltonian~\eqref{H_2D}.
We believe our DTQW to be better in representing 2D physics in the sense that it exhibits dynamics of a 2D harmonic oscillator as we demonstrated in Fig.~\ref{orbit_1000}.

In the numerical calculation for Fig.~\ref{orbit_1000}, we set the system size to $L_x=L_y=101$ with $-50\le x\le50$ and $-50\le y\le50$ under periodic boundary conditions in both directions.
We used the effective potential of the form
\begin{align}\label{potential_x}
\theta_{x_\mu}(x_\mu)=
\begin{cases}
\pi/4 \quad  & \mbox{for $5<x_\mu\le 50$},\\
bx_\mu \quad & \mbox{for $\abs{x_\mu}\le 5$},\\
-\pi/4 \quad & \mbox{for $-50\le x_\mu<-5$},
\end{cases}
\end{align}
where $x_1=x$, $x_2=y$ and $b=\pi/20$.
We repeated numerical multiplication of $U^{(2)}$ to the initial state.
For the initial state, we used an eigenstate of the eigenvalue unity of the time-evolution operator $U^{(2)}$ shifted in the $x$ direction by two sites and imposed the initial velocity in the form of $\ee^{\ii(k_xx+k_yy)}$ with $(k_x, k_y)=(0, \pi)$.
The eigenstate of the eigenvalue unity of $U^{(2)}$ is in the Trotter limit given by the Gaussian form of the zero-energy eigenvalue of our extended 2D Dirac Hamiltonian~\eqref{H_2D}, which we explicitly obtain in App.~\ref{appendeigstate-linearpot-2D}.

\begin{figure}
  \centering
\includegraphics[width=0.5\columnwidth]{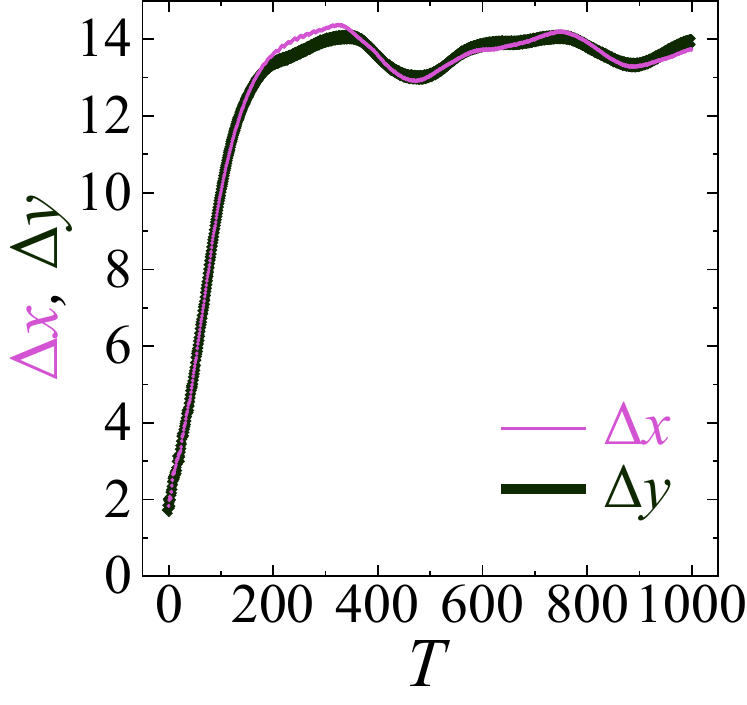}
  \caption{(Color online). The time-step dependence of the standard deviations $\Delta x$ (thin purple line) and $\Delta y$ (thick dark green line) of the quantum walker in 2D. We set $\hbar=a=\Delta t=1$.}
  \label{SD_1000}
\end{figure}

Figure~\ref{orbit_1000} shows the expectation values, 
\begin{align}
\mqty{
\expval{x(T)}:=\sum_{x, y}xP(x, y, T), \\
\expval{y(T)}:=\sum_{x, y}yP(x, y, T),
}
\end{align}
at each time step, where $P(x, y, T)$ is the quantum probability at site $(x, y)$ at time step $T$ and satisfies $\sum_{x, y}P(x, y, T)=1$.
We observe the circular trajectory in Fig.~\ref{orbit_1000}; after some time it converges to an orbit of a limit cycle (Fig.~\ref{orbit_1000}(b)), which closely resembles the one of a Schr\"{o}dinger dynamics under a 2D harmonic potential.
In Fig.~\ref{SD_1000}, we can see that the standard deviations almost converge to a constant after $T\sim 500$, which implies that the walker reaches a steady state of a circling wave packet around the time.
With these facts, we believe that we successfully observe dynamics that resembles the 2D harmonic oscillator.
We confirm in App.~\ref{appendeigstate-linearpot-2D} that all the eigenstates of the corresponding 2D extended Dirac Hamiltonian~\eqref{H_2D} are composed of the eigenstates of a 2D harmonic oscillator.

The fact that the present 2D DTQW behaves like a 2D harmonic oscillator is particularly important to some of the present authors for studies of quantum active matter.
They defined in Ref.~\cite{Yamagishi23} a quantum version of the active Brownian particle~\cite{Schweitzer98}, in which Schweitzer \textit{et al.}\ numerically demonstrated that a classical active particle climbs up the 2D harmonic potential and makes a circular orbit.
Some of the present authors~\cite{Yamagishi23} are reproducing similar movement of the quantum version, using the present oscillator behavior of the 2D DTQW.
This is why the present quantum walker's making the circular orbit is critically important.

\subsection{\label{sec2B}Topological Edge States of Two-Dimensional DTQW}
Let us turn to topological properties of DTQW.
Jackiw and Rebbi~\cite{Jackiw76} suggested that 
when a Dirac system 
e.g.\ Eq.~\eqref{H_1D} 
(presumed to be extended to infinity)
has two domains
in each of which
the mass term takes a different value e.g.,
\begin{align}\label{domainwall}
m_x(x)=\begin{cases}
m_1\quad&\mbox{for $x>0$}, \\
m_2\quad&\mbox{for $x<0$},
\end{cases}
\end{align}
then
a robust zero-energy state spatially localized in the vicinity of the domain wall
emerges in the mass gap
iff
the sign of $m_1$ and $m_2$ differ~%
\cite{Shen13}.
In other words,
the zero-energy domain-wall state is protected by 
an index $\nu_1$
defined as
$(-1)^{\nu_1}={\rm sgn}(m_1){\rm sgn}(m_2)$,
which takes two integral values $\nu_1=0,1$ in this particular case.
Now that
the concept of topological insulator is well established,
Jackiw and Rebbi's example is recognized as its earliest
realization, and the index $\nu_1$ is 
interpreted as
a topological number.
Indeed, the model~\eqref{H_1D} belongs to the symmetry class DIII
with the topology of type $\mathbb{Z}_2$ in 1D~\cite{Kitaev09, Ryu10}.

We checked it numerically
using the 1D DTQW prescribed by Eqs.~\eqref{eq10} and~\eqref{eq11}
with $\theta_x(x)$ having two domains, 
\begin{align}\label{wallx}
\theta_x(x)=\begin{cases}
\theta_1 \quad & \mbox{for $|x|<L_1$},\\ 
\theta_2 \quad & \mbox{for $L_1<|x|<[L_x/2]$}.
\end{cases}
\end{align}
Note that each of the domain walls at $x=\pm L_1$ corresponds to the one in Eq.~\eqref{domainwall} as in $\theta_x(x)=m_x(x)\Delta t$.
Since our DTQW is under the periodic boundary condition,
there are 
two domain walls with discontinuities in $m_x(x)$ at $x=\pm L_1$, 
and therefore we observed two 
topologically 
protected zero-energy states, 
each of which is localized at a different domain wall.
(Incidentally, squaring the Dirac Hamiltonian as in Eq.~\eqref{eq13}, 
we find that
the last term of $H_{\sch}^{(1)}$ yields delta functions at the discontinuities of $m_x(x)$, 
and thus
the edge states of the Dirac Hamiltonian can also be interpreted as 
bound states of 
the corresponding Schr\"{o}dinger particle to the delta potentials.)

In the 2D realization of our DTQW
prescribed by Eqs.~\eqref{eq22} and~\eqref{eq23},
using again
the domain-wall configuration of $\theta_x(x)$ introduced in the 1D case with $\theta_y(y)\equiv 0$ in Eq.~\eqref{eq22},
the protected zero-energy states acquire a dispersion; see App.~\ref{appendedge2D} for the solution in the case of the 2D extended Dirac Hamiltonian~\eqref{H_2D}.
Figure~\ref{bands}(a) shows quasi-energy spectrum $E_n=-\ii\log{U^{(2)}_n}$ for each ${k_y}_n=2n\pi/L_y$ with $\theta_y=0$.
We observe that two linear dispersions with positive and negative slopes
completely traverses the bulk energy gap, manifesting the feature of protected gapless edge states.
Their gaplessness is protected by a topological number $\nu_1=1$
introduced above;
when $\nu_1$ changes upon changing $\theta_x(x)$, the bulk energy gap must close once and reopen
in the space of control parameters, 
where different topological phases are defined.

The extended Dirac Hamiltonian \eqref{H_2D} with $m_y=0$ has a time-reversal symmetry under $\Theta=\sigma^x\otimes\tau^yK$ with $K$ being complex conjugation, a particle-hole symmetry under $\Xi=\mathbbm{I}_{4\times4}K$, and a chiral symmetry under $\Pi=\sigma^x\otimes\tau^y$,
and hence belongs to the symmetry class DIII with a topology 
of type 
$\mathbbm{Z}_2$ in 2D~\cite{Kitaev09, Ryu10}.
However, the time-evolution operator $U^{(2)}=S_yS_xC_x$ (with $\theta_y=0$) only has the particle-hole symmetry under $\Xi=\mathbbm{I}_{4\times4}K$ because of the specific ordering of $S_yS_x$ ($\neq S_xS_y$), and hence our DTQW belongs to the symmetry class D with a topology
of type
$\mathbbm{Z}$ in 2D~\cite{Kitaev09, Ryu10}.
(Incidentally, we have an additional sublattice symmetry in $S_y$. 
Adding the phase $\ee^{\ii\pi}$ to the every other $y$ and shifting $k_y$ with $\pi$ do not change $S_y$.
This results in a $\pi$-periodicity in $k_y$ in the spectra in Fig.~\ref{bands}.)

\begin{figure}
\centering
\includegraphics[width=0.85\columnwidth]{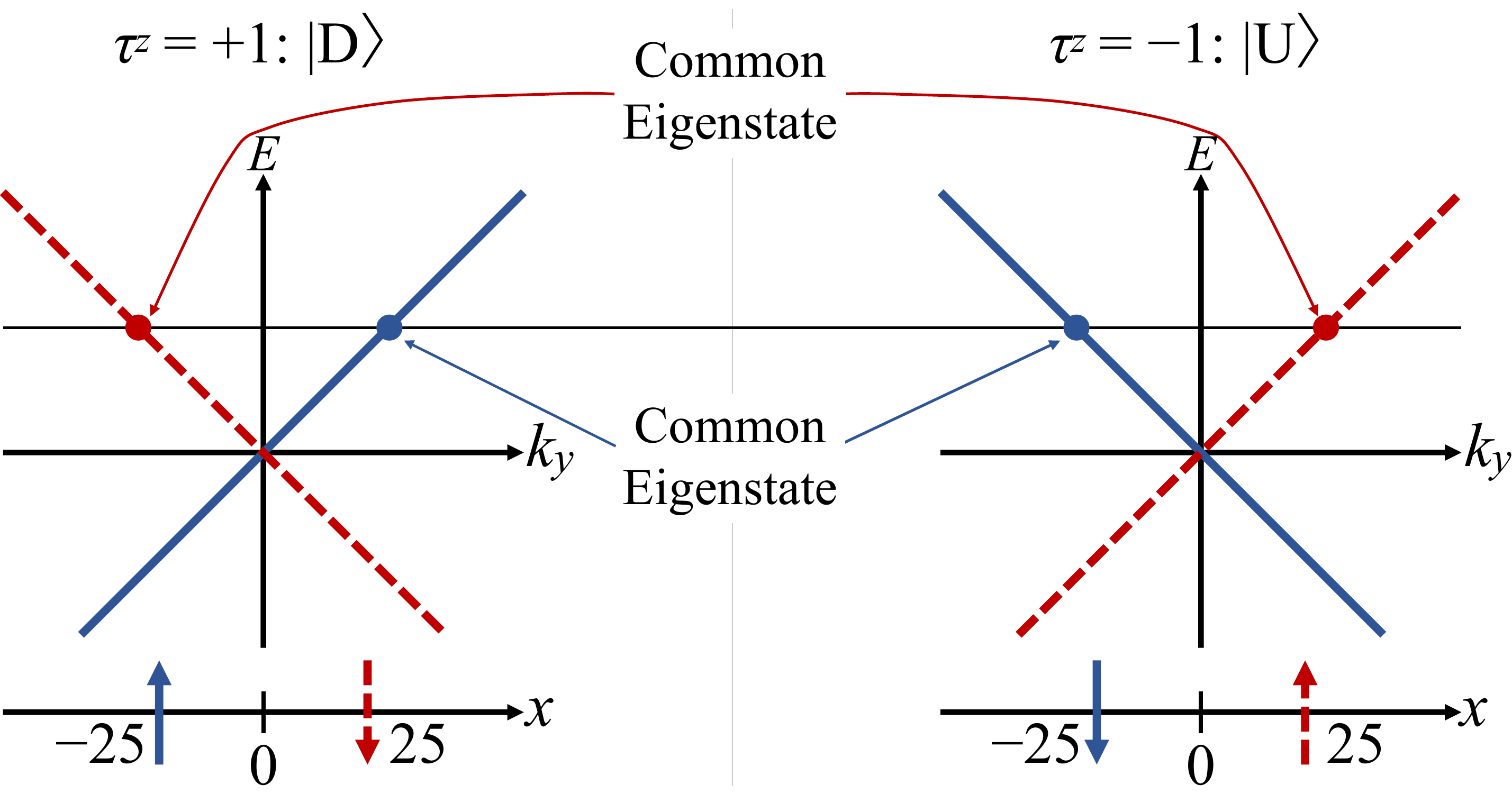}
\caption{Schematic representation of the dispersion around $k_y=0$ in Fig.~\ref{bands}(a). We set $\hbar=a=\Delta t=1$.}
\label{dispersion}
\end{figure}

\begin{figure}
\centering
\includegraphics[width=0.65\columnwidth]{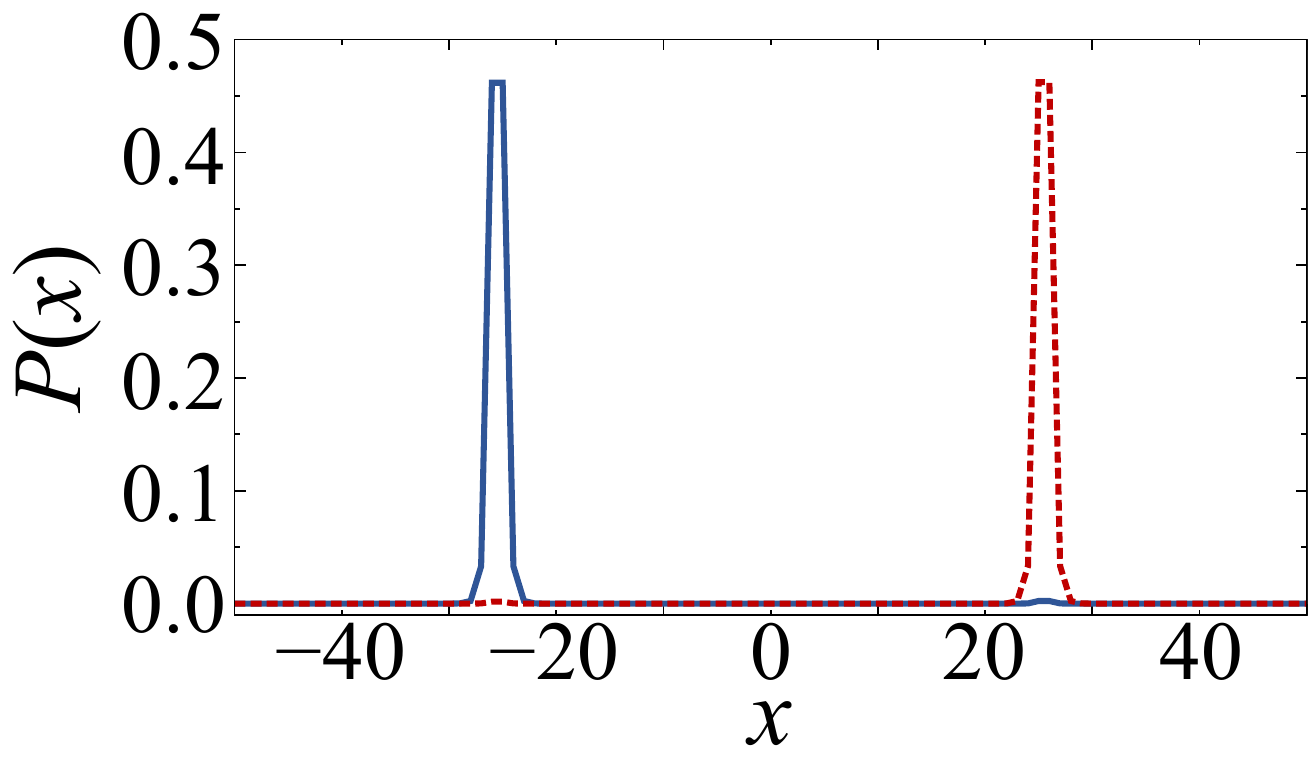}
\caption{Probability density of an edge state localized around $x=-25$ (solid blue line) and one localized around $x=25$ (red broken line) with $k_y=0$ and the zero quasi-energy in the spectrum of Fig.~\ref{bands}(a). We set $\hbar=a=\Delta t=1$.}
\label{localized}
\end{figure}

We can understand the structure of the dispersion of edge states in Fig.~\ref{bands}(a) as in Fig.~\ref{dispersion}.
Let us first note that our time-evolution operator $U^{(2)}=S_yS_xC_x$ for $\theta_y=0$ is block-diagonalized for the blocks $\tau^z=\pm1$ with the same absolute value of $k_y$ but with a different sign.
Since each block belongs to the class D, we have a topology of type $2\mathbbm{Z}$.
In the block of $\tau^z=1$, namely $\ket{\mathrm{D}}$, 
an edge state localized at $x=-L_1=-25$ in Fig.~\ref{localized} 
has the dispersion of a positive slope and one at $x=L_1=25$ 
has one with a negative slope  as shown in Fig.~\ref{dispersion}.
In the block of $\tau^z=-1$, namely $\ket{\mathrm{U}}$, on the other hand,
an edge state at $x=-L_1=-25$ has the dispersion with a negative slope and the other at $x=L_1$ 
has one with a positive slope.
Since the two blocks have opposite signs of $k_y$, the dispersion has a mirror symmetry, and hence the eigenstates on the two solid lines in Fig.~\ref{dispersion} are common to each other; the same applies to the eigenstates on the broken lines.
This is why we observe the lines with both positive and negative slope crossing at $k_y=0$.
Two eigenvalues are degenerate on each line.

Upon introducing a nonzero value of $\theta_y$, which is incompatible with the dictated symmetry of class D, a gap emerges around $k_y=0$ as shown in Fig.~\ref{bands}(b).
Meanwhile, the topological edge states are robust against other types of small perturbation, which 
we numerically confirm by introducing randomness.
We added to $\theta_x(x)$
a random perturbation $\Delta{\theta_x}_0(x)$, randomly choosing  independently for each site uniformly from the range $[-0.25, 0.25]$.
As we see in Fig.~\ref{bands}(c), the degeneracy for $\tau^z=\pm 1$ is lifted but the crossing at $k_y=0$ remains.

\subsection{\label{sec2C}Chiral Symmetry and Higher-Order Topology}
In Fig.~\ref{bands}(b) and in its description,
we saw that
the presence 
of a finite value of $\theta_y$ is
incompatible with the symmetry dictated in the periodic table 
for the symmetry class D,
and hence
the edge states protected by
the standard first-order topology
have been gapped out.
However,
we now see that
the chiral symmetry
inherent to the 1D Dirac Hamiltonian~\eqref{H_1D}
leads to the emergence of the so-called
higher-order topology~
\cite{BBH,Hayashi18,Schindler,trif,ZF,BBH2},
which is beyond the standard classification of topological insulators
dictated by the periodic table given in Refs.~\cite{Kitaev09, Ryu10}.

The standard topological insulator
is characterized by the existence of
protected gapless or zero-energy surface states.
In $d$ space dimensions,
such surface states
appear on $(d-1)$-dimensional surfaces of the system.
In the case of the recently proposed 
higher-order topological insulator~\cite{BBH,Hayashi18,Schindler,trif,ZF,BBH2},
not only the $d$-dimensional bulk 
but also the $(d-1)$-dimensional surfaces
are both gapped, and yet,
some higher-order, e.g.\ $(d-n)$-dimensional ``surfaces''
(an extremity of the system with co-dimension $n$)
remain gapless with $n \ge 2$.
To represent such a higher-order surface,
the word ``corner'' is most commonly employed,
which in the case of $d=2$ and $n=2$ as in the present case
is consistent with the common usage of the word, as we will see below.

Let us 
note that the Pauli matrix $\sigma^x$ introduced along with $H_{\dirac_y}$ in Eq.~\eqref{H_2D}
is nothing but the chiral operator, i.e.\ $\Gamma_1=\sigma^x$, 
associated with the 1D Dirac Hamiltonian $H_{\dirac_x}$:
\begin{equation}
\{\Gamma_1, H_{\dirac_x}\}=0
\quad\mbox{with}\quad{\Gamma_1}^2=1.
\label{eqchiral}
\end{equation}
This being said, we notice that
the construction of the 2D extended Dirac Hamiltonian in
Eq.~\eqref{H_2D}
is done precisely in the same manner
as in the recipe in Ref.~\cite{Hayashi18} 
for constructing
the second- and higher-order ($n$th-order) topological insulators, 
starting with the
standard first-order topological insulators 
$H_1$ and $H_2$ 
as its building blocks, where
$H_1$ must have the chiral symmetry $\Gamma_1$ as in 
$\{\Gamma_1, H_1\}=0$ with ${\Gamma_1}^2=1$.
One can indeed show that the Hamiltonian $H^{(2)}$
constructed as
\begin{equation}
H^{(2)}= H_1\otimes 1 + \Gamma_1\otimes H_2
\label{H2}
\end{equation}
has the designed property of the second-order topological insulator~\cite{Hayashi18}.
Higher-order ($n$th-order) topological insulators $H^{(n)}$ are constructed with $n-1$ chiral operators and $n$ Hamiltonian of which at least $n-1$ anticommute with the corresponding chiral operators:
\begin{align}\label{Hn}
H^{(n)}=
&\underbrace{H_1\otimes1\otimes1\otimes\cdots\otimes1\otimes1}_{n}+%
\underbrace{\Gamma_1\otimes H_2\otimes1\cdots\otimes1\otimes1}_{n} \notag \\
&+\cdots+%
\underbrace{\Gamma_1\otimes\Gamma_2\otimes\Gamma_3\otimes\cdots\otimes\Gamma_{n-1}\otimes H_n}_{n}
\end{align}
with
\begin{align}\label{Gamman}
\left\{\Gamma_i, H_i\right\}=0,\quad{\Gamma_i}^2=1,\quad i=1,\cdots,n-1.
\end{align}

In the present case of our 2D extended Dirac Hamiltonian~\eqref{H_2D},
we can naturally identify the constituents as
$H_1=H_{\dirac_x}$ and 
$H_2=H_{\dirac_y}$.
Appendix~\ref{appendchiral} shows that the zero-energy eigenstate of our 2D extended Dirac Hamiltonian is the product of the
zero-energy edge state running in the $y$ direction and that running in the $x$ direction,
which results in the corner states demonstrated below in Fig.~\ref{corner} for our 2D DTQW model.
%
Surprisingly, our 3D extended Dirac Hamiltonian~\eqref{eq31} naturally satisfies the conditions~\eqref{Hn} and \eqref{Gamman} under the identification of $H_1=H_{\dirac_x}$, $H_2=H_{\dirac_y}$ and $H_3=H_{\dirac_z}$.
We can naturally apply the same argument to the 3D case as in the 2D case.

The appearance or non-appearance of a higher-order topological state 
(specifically a zero-energy corner state in the case of $n=2$)
is encoded in a topological index $\nu^{(n)}$
expressed 
(at least for a corner with a right angle~\cite{yoshi,trif})
as a product of 
conventional topological indices 
\begin{align}
\nu^{(n)}=\prod_{m=1}^n \nu_{m},
\end{align}
where each $\nu_{m}$ 
provides information on the existence and the absence of a gapless $(d-1)$-dimensional surface state
of the constituent first-order topological insulators $H_m$ in $d$ dimensions.
%

Specifically for $n=2$ and $d=2$ in the present case, as each of the two indices $\nu_{1}$ and $\nu_{2}$
encodes information on the existence and the absence of a gapless one-dimensional surface state,
the situation $\nu_{1},\nu_{2} =0$ 
corresponds to the absence, indicating that the system is trivial,
while
the situation $\nu_{1},\nu_{2} \neq 0$ signifies that
the system is topologically non-trivial,
so that
$\nu^{(2)}$ encodes
information on the existence and absence of 
a gapless zero-dimensional corner state.

In order to let corner states emerge in our 2D DTQW model,
we introduce the domain structure in the $x$ direction
also in the $y$ direction; 
we set $\theta_y(y)$ such that
\begin{align}\label{wally}
\theta_y(y)=
\begin{cases}
\theta_1 \quad & \mbox{for $|y|<L_2$},\\
\theta_2 \quad & \mbox{for $L_2<|y|<[L_y/2]$}
\end{cases}
\end{align}
in addition to the one in Eq.~\eqref{wallx}.
We chose the parameter values specifically as
$\theta_1=-\theta_2=\pi/3$, $L_x=L_y=101$ and 
$L_1=L_2=25$ for numerical calculation for Fig.~\ref{corner}.
We can observe four zero-energy corner states
localized at the four corners of the domain $|x|<L_1$ with $|y|<L_2$.
\begin{figure}
  \centering
  \includegraphics[width=0.58\columnwidth]{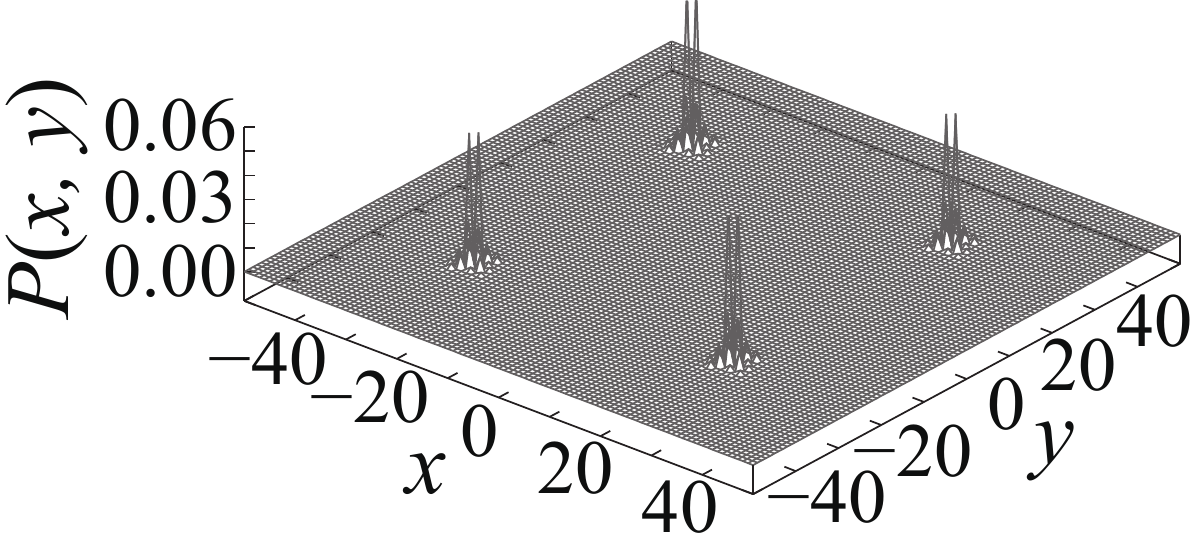}
  \caption{Probability distribution of zero-energy corner states with the potential $\theta_x(x)$ and $\theta_y(y)$ specified in the main text. 
We here plot one out of the totally eight corner states that are degenerate to a nearly zero eigenvalue. We set $\hbar=a=\Delta t=1$.}
\label{corner}
\end{figure}
Note that each corner state is defined as localized at one of the four corners of the domain;
the state represented in Fig.~\ref{corner} is a superposition of the four corner states.


\section{\label{summary}Summary}
To summarize, we proposed a new DTQW in multi-dimensional systems, whose continuum limit is the extended Dirac equation which can be further mapped to the Schr\"{o}dinger equation.
We successfully reproduced with our DTQW the dynamics similar to that of a Schr\"{o}dinger particle under a harmonic potential.
We also observed topological edge and corner states with discontinuous effective potentials in one- and two-dimensional systems simply by manipulating the coin operators of our DTQW.
We thereby claim that the present DTQW is a powerful platform of numerical simulation and experimental implementation of the Dirac and Schr\"{o}dinger particles.

\begin{figure}
  \centering
  \includegraphics[width=\columnwidth]{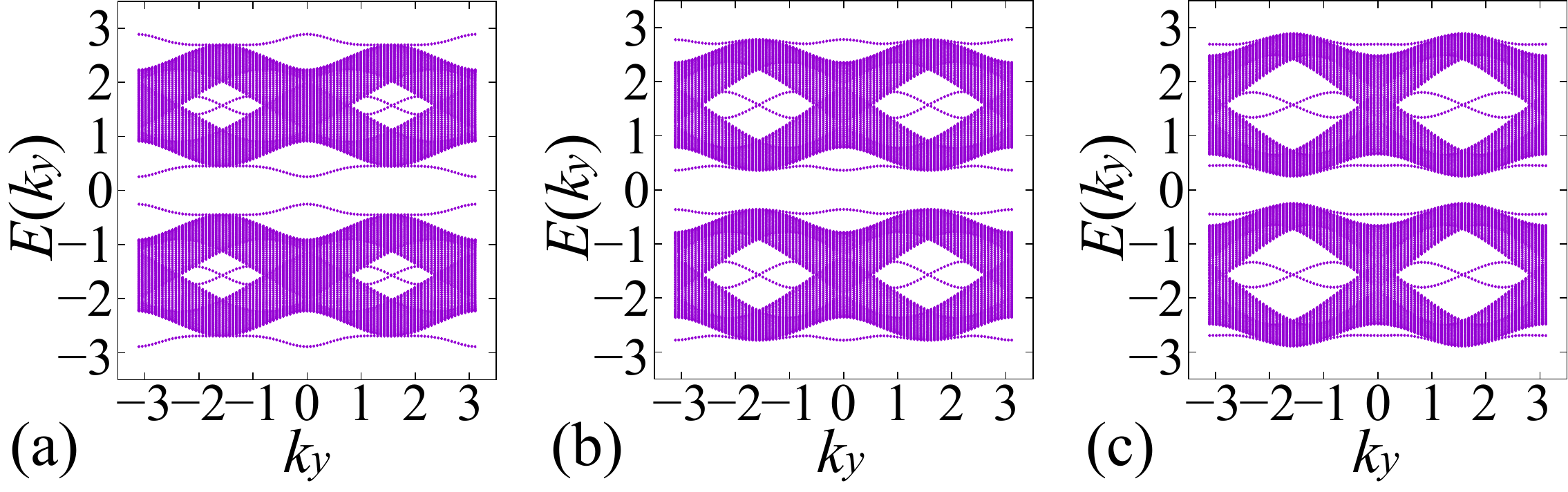}
\caption{Dispersion relation of the quasi-energy spectra $E(k_y)$ for the domain-wall structure in $\theta_x(x)$ as specified in the main text with (a) $\theta_y=\pi/6$, (b) $\theta_y=\pi/4$ and (c) $\theta_y=\pi/3$. We set $\hbar=a=\Delta t=1$.}
\label{hijacked}
\end{figure}

As a final remark, increasing $\theta_y$ further from the case in Fig.~\ref{bands}(b), we find the spectrum in Fig.~\ref{hijacked}.
We have numerically confirmed that the states enclosed in the openings of the bulk bands are edge states; see Appendix~\ref{appendmomotaro} of the bulk band structure.
This implies that our DTQW accommodates a further symmetry that protects these enclosed edge states, but we have not resolved yet what symmetry it is. 

\begin{acknowledgments}
It is a pleasure to acknowledge discussion about topological properties with Dr. Franco Nori.
This work is supported by JSPS KAKENHI Grant Numbers JP19H00658, JP20H01828, JP20K03788, JP21H01005, and JP22H01140.
\end{acknowledgments}

\appendix

\section{\label{appendeigstate}Eigenvalues of the 2D extended Dirac Hamiltonian}

In the present Appendix, we describe how to obtain the eigenvalues and eigenvectors of our 2D extended Dirac Hamiltonian~\eqref{H_2D} out of those of the 1D Dirac Hamiltonian~\eqref{H_1D}.
The former should give the quasi-energy eigenvalues and eigenvectors of our 2D DTQW within the Trotter approximation, particularly near zero energy, that is, near the eigenvalue unity of the time-evolution operator.

We first introduce the general formalism in App.~\ref{appendeigstate0}.
We then present the explicit solutions in the case of linear mass terms in App.~\ref{appendeigstate-linearpot} and in the case of stepwise mass terms in App.~\ref{appendtopo}.

\subsection{\label{appendeigstate0}General formalism}

We first set the  eigenstates of $H_{\dirac_x}$ $H_{\dirac_y}$  as in
\begin{align}\label{sol1Dx}
H_{\dirac_x}\ket*{\psi^{\dirac,(1)}_{E_x}}&=E_x\ket*{\psi^{\dirac,(1)}_{E_x}},
\\\label{sol1Dy}
H_{\dirac_y}\ket*{\psi^{\dirac,(1)}_{E_y}}&=E_y\ket*{\psi^{\dirac,(1)}_{E_y}}
\end{align}
with the normalization
\begin{align}
\braket*{\psi^{\dirac,(1)}_{E_x}}=\braket*{\psi^{\dirac,(1)}_{E_y}}=1.
\end{align}
For a shorthand, let their direct product denoted by
\begin{align}
\ket{\psi^{\dirac,(1)}_{Ex,E_y}}:=\ket*{\psi^{\dirac,(1)}_{E_x}}\ket*{\psi^{\dirac,(1)}_{E_y}}.
\end{align}

We now assume the Ansatz for the eigenstate of our 2D extended Dirac Hamiltonian
\begin{align}\label{H_2D-1}
&H_{\dirac}^{(2)}
:=H_{\dirac_x}\otimes\tau^0+\sigma^x\otimes H_{\dirac_y} 
\end{align}
of the form
\begin{align}\label{eigstate2D}
\ket{\psi^{\dirac,(2)}_{E}}=\qty(\gamma+\delta\sigma^x)\ket{\psi^{\dirac,(1)}_{Ex,E_y}},
\end{align}
where $\gamma$ and $\delta$ are real coefficients to be determined hereafter.
From the eigenvalue equation
\begin{align}
H_{\dirac}^{(2)}\ket{\psi^{\dirac,(2)}_{E}}=E\ket{\psi^{\dirac,(2)}_{E}},
\end{align}
we obtain
\begin{align}
&\qty[\gamma\qty( E_x + E_y \sigma^x )+\delta\qty(-E_x\sigma^x+E_y)] \ket{\psi^{\dirac,(1)}_{Ex,E_y}}
\notag\\
&\qquad=E\qty(\gamma+\delta\sigma^x)\ket{\psi^{\dirac,(1)}_{Ex,E_y}},
\end{align}
where we used the anti-commutation relation
\begin{align}
\acomm{H_{\dirac_x}}{\sigma^x}=\acomm{\qty(\epsilon\sigma^zp_x+m_x(x)\sigma^y)}{\sigma^x}=0.
\end{align}
We thereby find the equations for the coefficients as 
\begin{align}\label{alpha-beta-1}
\gamma E_x +\delta E_y&=\gamma E,
\\\label{alpha-beta-2}
\gamma E_y-\delta E_x &=\delta E.
\end{align}

First, let us eliminate $E$ from the set of the equations.
We then find
\begin{align}\label{eqA12}
2\gamma\delta E_x=\qty(\gamma^2-\delta^2)E_y,
\end{align}
which motivates us to define the transformation of the coefficients of the forms
\begin{align}\label{eqA13}
\gamma=A\cos\phi,\quad\delta=A\sin\phi.
\end{align}
We then have from Eq.~\eqref{eqA12}
\begin{align}\label{eqA14}
\tan2\phi=\frac{E_y}{E_x},
\end{align}
which determines the phase coefficient $\phi$ for the specific solutions of Eqs.~\eqref{sol1Dx} and~\eqref{sol1Dy}.
The amplitude coefficient $A$, on the other hand, is found from the normalization
\begin{align}\label{eqA15}
1&=\braket{\psi^{\dirac,(2)}_{E}}=\gamma^2+\delta^2+2\gamma\delta s
\notag\\
&=A^2(1+s\sin2\phi),
\end{align}
where $s:=\ev{\sigma^x}{\psi^{\dirac,(1)}_{E_x}}$.
We let $A$ undetermined in the present Appendix since it depends on the specific form of the eigenstate $\ket{\psi^{\dirac,(1)}_{E_x}}$.

The set of equations~\eqref{alpha-beta-1} and~\eqref{alpha-beta-2} further produces
\begin{align}
E_x=E\cos2\phi,\qquad
E_y=E\sin2\phi,
\end{align}
and hence
\begin{align}\label{totE}
E=\pm\sqrt{{E_x}^2+{E_y}^2}.
\end{align}
This implies that  our 2D extended Dirac Hamiltonian~\eqref{H_2D} is indeed a precise direct product of independent components of 1D Dirac Hamiltonians $H_{\dirac_x}$ and $H_{\dirac y}$, and further implies that the 2D DTQW presented in Sec.~\ref{sec2}  is also a precise direct product of independent components of 1D DTQW in the $x$ and $y$ directions

From Eq.~\eqref{totE} we can conclude the following.
First, the zero-energy eigenstate of $H_{\dirac}^{(2)}$, if any, can be constructed only from the zero-energy eigenstates of $H_{\dirac_x}$ and $H_{\dirac_y}$, which is indeed simply given by 
\begin{align}\label{zeroeig}
\ket{\psi^{\dirac,(2)}_{E=0}}=\ket{\psi^{\dirac,(1)}_{E_x=0}}\ket{\psi^{\dirac,(1)}_{E_y=0}}.
\end{align}
Second, if there is an energy gap  in the spectrum of either of $H_{\dirac_x}$ or $H_{\dirac_y}$, then the spectrum of $H_{\dirac}^{(2)}$ has an energy gap.

\subsection{\label{appendeigstate-linearpot}Case of the linear potentials~\eqref{linearpot0}}

We here explicitly obtain the Gaussian form of the zero-energy eigenstate of the 2D extended Dirac Hamiltonian~\eqref{H_2D} under the linear potentials~\eqref{linearpot0}.
The eigenstate of the eigenvalue unity of the time-evolution operator $U^{(2)}$, which state we used for the initial state of our simulation in Subsec.~\ref{sec2A}, is given by the state given here within the Trotter approximation.

\subsubsection{\label{appendeigstate-linearpot-1D}Eigenvalues of the 1D Dirac Hamiltonian~\eqref{H_1D}}
Let us first derive eigenstates of the 1D Dirac Hamiltonian~\eqref{H_1D} with $m_x(x)=\beta x$.
The Schr\"{o}dinger Hamiltonian after the unitary transformation $V=\exp(\ii\sigma^y\pi/4)$ reads
\begin{align}\label{eqB12}
\tilde{H}_{\sch}^{(1)}%
&:=\mqty(
\tilde{H}_{\sch_+}^{(1)} & 0 \\
0 & \tilde{H}_{\sch_-}^{(1)}
),
\end{align}
where $\tilde{H}_{\sch_{\pm}}^{(1)}:=\epsilon^2{p_x}^2+\beta^2x^2\pm\epsilon\beta$ as in Eq.~\eqref{sch1}.
Each of the block Hamiltonians can be rewritten in the form 
\begin{align}
\tilde{H}_{\sch_{\pm}}^{(1)}=\omega\qty(\hat{a}^\dag\hat{a}+\frac{1}{2}\pm\frac{1}{2})
\end{align}
with the ladder operators
\begin{align}\label{eqB11}
&\hat{a}^\dag=\displaystyle\frac{1}{\sqrt{2}}\qty(-\sqrt{\frac{\epsilon}{\beta}}\dv{x}+\sqrt{\frac{\beta}{\epsilon}}x), \notag \\
&\hat{a}=\displaystyle\frac{1}{\sqrt{2}}\qty(\sqrt{\frac{\epsilon}{\beta}}\dv{x}+\sqrt{\frac{\beta}{\epsilon}}x),
\end{align}
where we employed the same identification for $\epsilon$, $\beta$ and $\omega$ as in Eq.~\eqref{eq170}.
Therefore, the Hamiltonian~\eqref{eqB12} is rewritten as follows:
\begin{align}\label{eqB12-1}
\tilde{H}_{\sch}^{(1)}%
&=\omega\qty(\hat{a}^\dag\hat{a}+\frac{1}{2}+\frac{1}{2}\sigma^z)
=\omega\mqty(
\hat{a}^\dag\hat{a}+1 & 0 \\
0 & \hat{a}^\dag\hat{a}
).
\end{align}

We thus find for the Hamiltonian~\eqref{eqB12-1}
that the following two eigenstates are degenerate in the energy eigenvalue $n\omega$ with $n>0$:
\begin{align}\label{eqB14}
\ket{\tilde{\psi}_{{E_x}_1}^{\sch, (1)}}=\mqty(
\ket{n-1} \\
0
)\quad\mbox{and}\quad\ket{\tilde{\psi}_{{E_x}_2}^{\sch, (1)}}=\mqty(
0 \\
\ket{n}
).
\end{align}
On the other hand, the eigenstate of the zero-energy eigenvalue is uniquely given by
\begin{align}\label{eqB15}
\ket{\tilde{\psi}_{E_x=0}^{\sch, (1)}}=\mqty(
0 \\
\ket{0}
).
\end{align}
We then obtain the eigenstates of energy eigenvalue $n\omega$ (with $n>0$) of the Hamiltonian $H_{\sch}^{(1)}$ defined in Eq.~\eqref{eq13} by a unitary transformation $V^\dag=\exp(-\ii\sigma^y\pi/4)$ as arbitrary superpositions of the following two states:
\begin{align}\label{eqB16}
\ket{\psi_{{E_x}_1}^{\sch, (1)}}=\frac{1}{\sqrt{2}}\mqty(
\ket{n-1} \\
\ket{n-1}
)\quad\mbox{and}\quad\ket{\psi_{{E_x}_2}^{\sch, (1)}}=\frac{1}{\sqrt{2}}\mqty(
-\ket{n} \\
\ket{n}
).
\end{align}
The unitary transformation above, upon applied to Eq.~\eqref{eqB15}, gives the eigenstate of the zero-energy eigenvalue of the Hamiltonian~\eqref{eq13} as follows:
\begin{align}\label{eqB17}
\ket{\psi_{E_x=0}^{\sch, (1)}}=\frac{1}{\sqrt{2}}\mqty(
-\ket{0} \\
\ket{0}
).
\end{align}
More specifically, the eigenfunction of the zero-energy eigenvalue takes the Gaussian form
\begin{align}
\psi_0^{(1)}(x)%
:=\ip{x}{\psi_{E_x=0}^{\sch, (1)}}%
=\qty(\frac{\beta}{4\pi\epsilon})^{1/4}\mqty(
-1 \\
1
)\exp[-\frac{\beta}{2\epsilon}x^2]
.
\end{align}

Since the Schr\"{o}dinger Hamiltonian $H_{\sch}^{(1)}$ in Eq.~\eqref{eq13} is the square of the 1D Dirac Hamiltonian~\eqref{H_1D}, we anticipate that the two degenerate eigenstates in the eigenvalue $n\omega$ of the former Hamiltonian split into the eigenstates of the eigenvalues $\pm\sqrt{n\omega}$ of the latter Hamiltonian.
In fact, with superposing the two states in Eq.~\eqref{eqB16}, we obtain a unique eigenstate of each of the energy eigenvalues $\sqrt{n\omega}$ and $-\sqrt{n\omega}$.
Since the eigenstates are superpositions of the two states in Eq.~\eqref{eqB16}, we first write the eigenstates as
\begin{align}\label{eqB18}
\frac{A}{\sqrt{2}}\mqty(
\ket{n-1} \\
\ket{n-1}
)+\frac{B}{\sqrt{2}}\mqty(
-\ket{n} \\
\ket{n}
)
\end{align}
and determine the relationship between the coefficients $A$ and $B$.

Let us operate the 1D Dirac Hamiltonian~\eqref{H_1D} from the left to the state above.
We can utilize the following expressions of $x$ and $p_x$ in terms of the ladder operators in Eq.~\eqref{eqB11}:
\begin{align}\label{eqB13}
x=\frac{1}{\sqrt{2}}\sqrt{\frac{\epsilon}{\beta}}(\hat{a}^\dag+\hat{a}),\quad%
p_x=\ii\frac{1}{\sqrt{2}}\sqrt{\frac{\beta}{\epsilon}}(\hat{a}^\dag-\hat{a}).
\end{align}
These expressions let us rewrite the 1D Dirac Hamiltonian~\eqref{H_1D} with its mass being $m_x(x)=\beta x$ as follows:
\begin{align}
H_\dirac^{(1)}%
&=\epsilon\sigma^zp_x+\beta x\sigma^y \notag \\
&=\mqty(
\epsilon p_x & -\ii\beta x \\
\ii\beta x & -\epsilon p_x
) \notag \\
&=
\frac{\sqrt{\omega}}{2}
\mqty(
-\ii(\hat{a}^\dag-\hat{a}) & -\ii(\hat{a}^\dag+\hat{a}) \\
\ii(\hat{a}^\dag+\hat{a}) & \ii(\hat{a}^\dag-\hat{a})
).
\end{align}
Hence operating the 1D Dirac Hamiltonian~\eqref{H_1D} to the state~\eqref{eqB18} from the left yields
\begin{align}
&H_\dirac^{(1)}\qty[
\frac{A}{\sqrt{2}}\mqty(
\ket{n-1} \\
\ket{n-1}
)+\frac{B}{\sqrt{2}}\mqty(
-\ket{n} \\
\ket{n}
)] \notag \\%
&=
\sqrt{n\omega}
\qty[
-\ii\frac{B}{\sqrt{2}}\mqty(
\ket{n-1} \\
\ket{n-1}
)+\ii\frac{A}{\sqrt{2}}\mqty(
-\ket{n} \\
\ket{n}
)].
\end{align}
We take $\ii A=\pm B$ and $\mp\ii B=A$ in order for the state~\eqref{eqB18} to be the eigenstates of the eigenvalues $\pm\sqrt{n\omega}$.

Thus, the eigenstate of the eigenvalue $+\sqrt{n\omega}$ of the 1D Dirac Hamiltonian~\eqref{H_1D} is uniquely given in the form of
\begin{align}\label{eqB15-1}
\ket{\psi_{E_x=+\sqrt{n}}^{\dirac, (1)}}=\frac{1}{\sqrt{2}}\mqty(
\ket{n-1} \\
\ket{n-1}
)+\frac{\ii}{\sqrt{2}}\mqty(
-\ket{n} \\
\ket{n}
),
\end{align}
while the eigenstate of the eigenvalue $-\sqrt{n\omega}$ of the 1D Dirac Hamiltonian~\eqref{H_1D} is uniquely given in the form of
\begin{align}\label{eqB16-1}
\ket{\psi_{E_x=-\sqrt{n}}^{\dirac, (1)}}=\frac{1}{\sqrt{2}}\mqty(
\ket{n-1} \\
\ket{n-1}
)-\frac{\ii}{\sqrt{2}}\mqty(
-\ket{n} \\
\ket{n}
).
\end{align}
Meanwhile, we can easily confirm that the eigenstate of the zero-energy eigenvalue of the Schr\"{o}dinger Hamiltonian~\eqref{eq13} is also the eigenstate of the zero-energy eigenvalue of the Dirac Hamiltonian~\eqref{H_1D} by operating it to the state~\eqref{eqB17}:
\begin{align}\label{eqB17-1}
\ket{\psi_{E_x=0}^{\dirac, (1)}}=\frac{1}{\sqrt{2}}\mqty(
-\ket{0} \\
\ket{0}
).
\end{align}

\subsubsection{\label{appendeigstate-linearpot-2D}Eigenvalues of the 2D extended Dirac Hamiltonian~\eqref{H_2D}}

We now construct the eigenstates of the 2D Dirac Hamiltonian~\eqref{H_2D} out of the eigenstates~\eqref{eqB15-1}--\eqref{eqB17-1} of the 1D Dirac Hamiltonian~\eqref{H_1D} following the general formalism presented in App.~\ref{appendeigstate0}.

Let us first find the state of the zero eigenvalue.
According to Eq.~\eqref{zeroeig} of the general formalism, the zero-energy eigenstate of the 2D Hamiltonian $H_{\dirac}^{(2)}$ is the direct product of the zero-energy eigenstates of the 1D Hamiltonians $H_{\dirac_x}$ and $H_{\dirac_y}$:
\begin{align}
\ket{\psi_{E_x=0}^{\dirac, (1)}}\ket{\psi_{E_y=0}^{\dirac, (1)}}=\frac{1}{2}\mqty(
-\ket{0} \\
\ket{0}
)\otimes \mqty(
-\ket{0} \\
\ket{0}
).
\end{align}
More specifically, the eigenfunction of the zero-energy eigenvalue takes the following 2D Gaussian form:
\begin{align}
\psi_0^{(2)}(x,y)&=\braket{x}{\psi_{E_x=0}^{\dirac, (1)}}\braket{y}{\psi_{E_y=0}^{\dirac, (1)}}
\notag\\
&=\sqrt{\frac{\beta}{4\pi\epsilon}}
\mqty(-1 \\ 1)\otimes\mqty(-1 \\ 1)
\exp\qty[-\frac{\beta}{2\epsilon}(x^2+y^2)].
\end{align}
This 2D Gaussian form is indeed close to what we used for the initial state in the numerical simulation of Fig.~\ref{orbit_1000}.

For non-zero eigenvalues, combinations of the 1D states with the eigenvalues $\pm\sqrt{m\omega}$ and with the eigenvalues $\pm\sqrt{n\omega}$ produce the 2D states with the eigenvalues $\pm\sqrt{(m+n)\omega}$ as in Eq.~\eqref{totE}.
The eigenstates are given in the form~\eqref{eigstate2D} with the coefficients specified by Eqs.~\eqref{eqA13}--\eqref{eqA15},
or more specifically by $\tan 2\phi=\pm\sqrt{n/m}$.
The degeneracy of the eigenvalues is the same as in the case of the 2D harmonic oscillator.

\subsection{\label{appendtopo}Topological edge states and higher-order topological corner state}

We finally show that our 2D extended Dirac Hamiltonian~\eqref{H_2D} with Eqs.~\eqref{wallx} and~\eqref{wally} has a zero-energy eigenstate in the product form of the zero-energy eigenstates of the 1D Dirac Hamiltonians~\eqref{H_1D} in the $x$ and $y$ directions, each of which under one of Eqs.~\eqref{wallx} and \eqref{wally}, respectively.
Since the zero-energy eigenstate of each of the latter Hamiltonians is an edge state, the zero-energy eigenstate of the former Hamiltonian is a corner state, being the product of edge states in different directions.

\subsubsection{\label{appendedge1D}First-order topological edge state in 1D}
We first solve the eigenvalue problem of the 1D Dirac Hamiltonian~\eqref{H_1D} 
\begin{align}\label{eqA37}
H_\dirac^{(1)}\ket{\psi_{E_x}^{\dirac, (1)}}%
&=(\epsilon\sigma^zp_x+m_x(x)\sigma^y)\ket{\psi_{E_x}^{\dirac, (1)}} \notag \\%
&=E_x\ket{\psi_{E_x}^{\dirac, (1)}},
\end{align}
where the mass term has a domain wall of the form
\begin{align}\label{wallxB}
m_x(x)=\begin{cases}
+m_0\quad\mbox{for $x>0$}, \\
-m_0\quad\mbox{for $x<0$}.
\end{cases}
\end{align}
We show that there is a bound state with the zero-energy eigenstate and scattering states with continuum spectra but with an energy gap.

Let us first focus on the bound state, assuming the zero-energy eigenvalue for explanatory purposes;
see Ref.~\cite{Shen13} for a solution without the assumption.
When we are focused on the zero-energy eigenvalue, 
the right-hand side of Eq.~\eqref{eqA37} vanishes, and therefore 
\begin{align}\label{eqA39}
(\epsilon\sigma^zp_x+m_x(x)\sigma^y)\ket{\psi_{E_x=0}^{\dirac, (1)}}=0.
\end{align}
For $x>0$, we have $m_x(x)=+m_0$, and hence the two rows of Eq.~\eqref{eqA39} respectively read
\begin{align}
-\ii\epsilon \dv{x}\psi_1 -\ii m_0 \psi_2&=0,
\\
\ii\epsilon \dv{x}\psi_2 +\ii m_0 \psi_1&=0,
\end{align}
where we used the notation
\begin{align}
\braket{x}{\psi_{E_x=0}^{\dirac, (1)}}=\mqty(\psi_1(x)\\ \psi_2(x)).
\end{align}
Since we obtain
\begin{align}\label{eqA43}
\dv[2]{x}\psi_1=\frac{m_0}{\epsilon}\dv{x}\psi_2=\qty(\frac{m_0}{\epsilon})^2\psi_1,
\end{align}
the convergent solution for $x>0$ is found to be
$\psi_1 \propto \psi_2 \propto \ee^{-\qty(m_0/\epsilon) x}$.
For $x<0$, instead of Eq.~\eqref{eqA43}, we have
\begin{align}
\dv[2]{x}\psi_1=-\frac{m_0}{\epsilon}\dv{x}\psi_2=\qty(\frac{m_0}{\epsilon})^2\psi_1,
\end{align}
and hence the convergent solution for $x<0$ is given by
$\psi_1 \propto \psi_2 \propto \ee^{+\qty(m_0/\epsilon) x}$.
To summarize after normalization, we obtain the eigenfunction of the zero-energy eigenvalue in the form of a bound state:
\begin{align}\label{edge1D}
\psi_0^{\dirac, (1)}(x)%
&:=\ip{x}{\psi_{E_x=0}^{\dirac, (1)}} \notag \\
&=\sqrt{\frac{m_0}{2\epsilon}}\mqty(
1 \\
1
)\ee^{-m_0|x|/\epsilon}.
\end{align}

Let us next find a scattering state, with an incoming wave proportional to $A\ee^{\ii k_x x}$ for $x<0$, a reflection wave proportional to $B\ee^{-\ii k_x x}$ for $x<0$, and a transmission wave proportional to $C\ee^{\ii k_x x}$ for $x>0$, where $k_x>0$.
For the incoming wave for $x<0$, the eigenvalue problem~\eqref{eqA37} reads
\begin{align}
\mqty(\epsilon k_y &\ii m_0 \\
-\ii m_0 & -\epsilon k_y)
\mqty(\psi_1 \\ \psi_2)
=E_x
\mqty(\psi_1 \\ \psi_2),
\end{align}
which yields
\begin{align}
\ket{\psi_{E_x}^{\dirac, (1)}}=
\begin{cases}
\mqty(\ii\cos\varphi \\ \sin\varphi)
\quad&\mbox{for $E_x=+\sqrt{\qty(\epsilon k_x)^2+{m_0}^2}$},
\\
\mqty(-\ii\sin\varphi \\ \cos\varphi)
\quad&\mbox{for $E_x=-\sqrt{\qty(\epsilon k_x)^2+{m_0}^2}$},
\end{cases}
\end{align}
where the coefficient $\varphi$ is defined in
\begin{align}
\tan 2\varphi=\frac{m_0}{\epsilon k_y}.
\end{align}

Let us hereafter focus on the scattering state with the positive energy eigenvalue $E_x=+\sqrt{(\epsilon k_x)^2+{m_0}^2}$.
We therefore assume the incoming wave of the form
\begin{align}\label{scat1D-A}
A\ee^{\ii k_x x}\mqty(\ii\cos\varphi \\ \sin\varphi)
\end{align}
for $x<0$.
Similarly, the reflection wave is given by
\begin{align}\label{scat1D-B}
B\ee^{-\ii k_x x}\mqty(\ii\sin\varphi \\ \cos\varphi)
\end{align}
for $x<0$, while the transmission wave is given by
\begin{align}\label{scat1D-C}
C\ee^{\ii k_x x}\mqty(-\ii\cos\varphi \\ \sin\varphi)
\end{align}
for $x>0$. 
In order for the first two on the left and the last on the right to be continuous at the origin, 
the amplitudes must satisfy
\begin{align}
A\cos\varphi+B\sin\varphi&=-C\cos\varphi,
\\
A\sin\varphi+B\cos\varphi&=C\sin\varphi.
\end{align}
They are followed by
\begin{align}
B&=-A\sin 2\varphi,
\\
C&=-A\cos 2\varphi,
\end{align}
which indeed satisfy the flux conservation $k_x\abs{A}^2=k_x\abs{B}^2+k_x\abs{C}^2$.
We can similarly find a solution for the negative energy eigenvalue  $E_x=-\sqrt{(\epsilon k_x)^2+{m_0}^2}$.

\begin{figure}
\centering
\includegraphics[width=0.9\columnwidth]{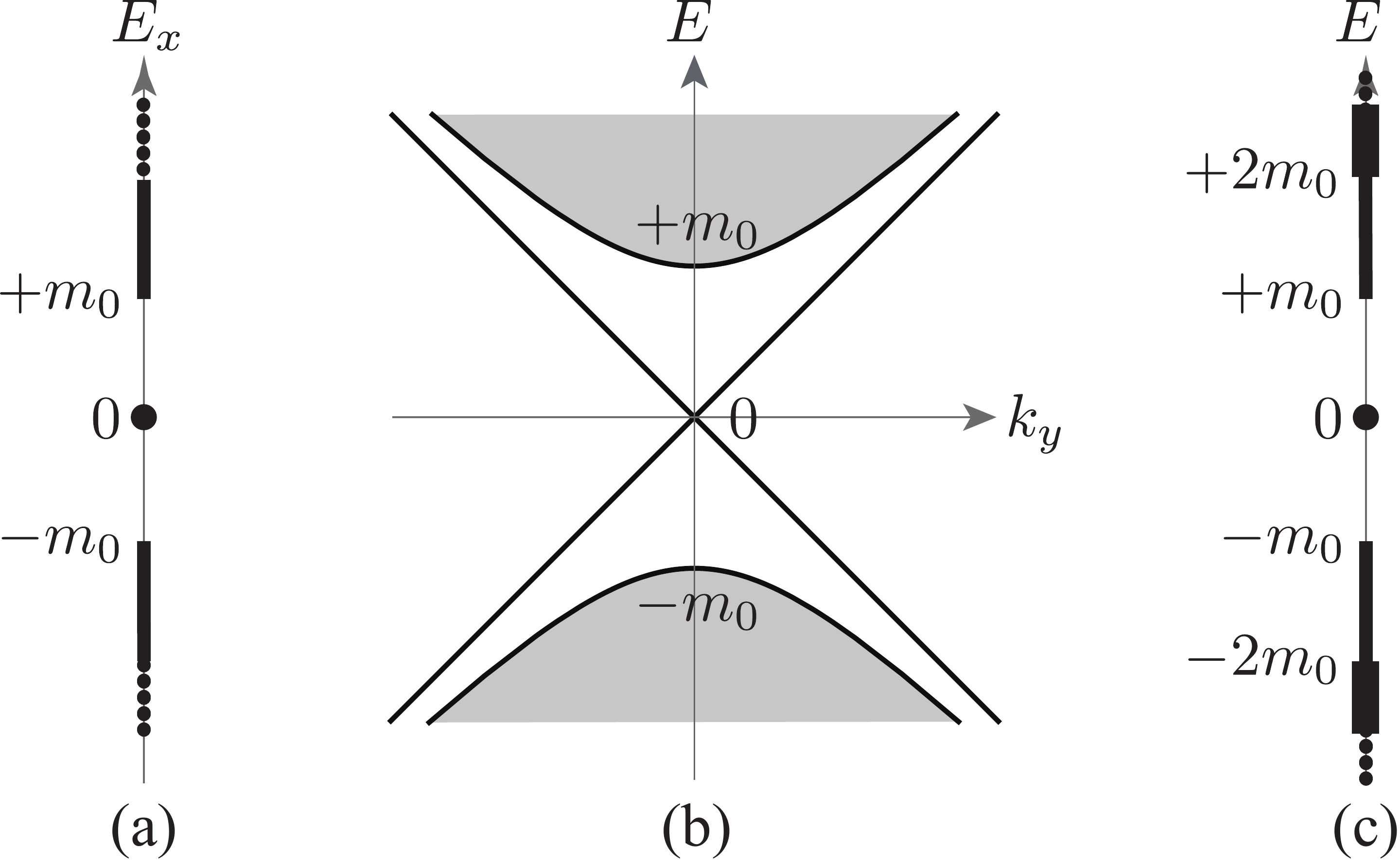}
\caption{(a) The energy spectrum of the 1D Dirac Hamiltonian with a domain wall in the mass term.
(b) The dispersion relation of our 2D extended Dirac Hamiltonian with a domain wall only in the mass term $m_x(x)$, whereas $m_y(y)\equiv0$.
(c) The energy spectrum of our 2D extended Dirac Hamiltonian with domain walls both in the mass term $m_x(x)$ and $m_y(y)$.}
\label{figA1}
\end{figure}

We thereby conclude that the scattering states have the energy continua of the forms $E_x=\pm\sqrt{(\epsilon k_x)^2+{m_0}^2}$ with the energy gap $-m_0<E_x<m_0$, in the middle of which exists the point-spectral bound state of the zero-energy eigenvalue; see
Fig.~\ref{figA1}(a).
This implies that the bound state is actually a topological edge state.

\subsubsection{\label{appendedge2D}First-order topological edge states in 2D}

We now find the eigenspectrum of our 2D extended Dirac Hamiltonian~\eqref{H_2D} but with $m_y(y)\equiv 0$.
This corresponds to the situation of 2D DTQW in Subsec.~\ref{sec2B}.

Following the general formulation in App.~\ref{appendeigstate0} again, we can find the eigenstates of $H_\dirac^{(2)}$ out of the eigenstates of $H_{\dirac_x}$ and $H_{\dirac_y}$. 
For  $H_{\dirac_x}$, we set the domain wall~\eqref{wallxB}, and hence its eigenstates are the ones given in App.~\ref{appendedge1D}, namely the topological edge state~\eqref{edge1D} with a point spectrum $E_x=0$ and the scattering states~\eqref{scat1D-A}--\eqref{scat1D-C} with the energy continuum $E_x=\pm\sqrt{(\epsilon k_x)^2+{m_0}^2}$.
For $H_{\dirac_y}$, on the other hand, we have only the kinetic term $\epsilon p_y \sigma_x\otimes\tau_z$, and therefore
the eigenvalues has the linear dispersions $E_y=\pm \epsilon k_y$ for the eigenvectors 
\begin{align}
\frac{1}{\sqrt{2}}\mqty(1 \\ \pm 1)\otimes\mqty(\ee^{\pm\ii k_y y} \\ 0)
\quad\mbox{and}\quad
\frac{1}{\sqrt{2}}\mqty(1 \\ \pm 1)\otimes\mqty(0 \\ \ee^{\pm\ii k_y y}).
\end{align}

We now combine the eigenvalues of the two Hamiltonians as in Eq.~\eqref{totE}.
Combining the point spectrum $E_x=0$ and the linear dispersions $E_y=\pm\epsilon k_y$ produce $E=\pm \epsilon k_y$.
Combining the continuum with the linear dispersions give 
\begin{align}\label{filledhyperbolic}
E=\pm\sqrt{(\epsilon k_x)^2+(\epsilon k_y)^2 + {m_0}^2}.
\end{align}
When we plot these eigenvalues for $k_y$, the former is the linear dispersions while the latter is hyperbolic curves filled by scanning $k_x$; see Fig.~\ref{figA1}(b).
This describes near the origin of the energy spectrum in Fig.~\ref{bands}(a) within the Trotter approximation.

\subsubsection{\label{appendchiral}Higher-order topological corner state in 2D}

We finally find the eigenspectrum of the 2D extended Dirac Hamiltonian~\eqref{H_2D} with domain walls both in $x$ and $y$ directions:
\begin{align}\label{wallyB}
m_y(y)=\begin{cases}
+m_0\quad\mbox{for $y>0$}, \\
-m_0\quad\mbox{for $y<0$}
\end{cases}
\end{align}
in addition to Eq.~\eqref{wallxB}.
This corresponds to the situation of 2D DTQW in Subsec.~\ref{sec2C}.

Both $H_{\dirac_x}$ and $H_{\dirac_y}$ now have the spectrum given in App.~\ref{appendedge1D}.
Combining them, we have the following for types of eigenvalues:
\setlength{\leftmargini}{24pt}
\setlength{\parskip}{0pt}
\setlength{\itemsep}{0pt}
\begin{enumerate}
\renewcommand{\labelenumi}{(\roman{enumi})}
\item $E_x=0$ and $E_y=0$ combine to produce $E=0$. This zero-energy eigenstate will be below identified as a second-order topological corner state.
\item $E_x=\pm\sqrt{(\epsilon k_x)^2+{m_0}^2}$ and $E_y=0$ combine to produce $E=\pm\sqrt{(\epsilon k_x)^2+{m_0}^2}$.
This is an edge state (the first-order topological state) in the $y$ direction, but a scattering state in the $x$ direction.
In other words, an edge continues in the $x$ direction, localized in the $y$ direction, on which a state propagates with the momentum $k_x$.
\item $E_x=0$ and $E_y=\pm\sqrt{(\epsilon k_y)^2+{m_0}^2}$ combine to produce $E=\pm\sqrt{(\epsilon k_y)^2+{m_0}^2}$.
In this case, an edge continues in the $y$ direction, localized in the $x$ direction, on which the state propagates with the momentum $k_y$.
\item $E_x=\pm\sqrt{(\epsilon k_x)^2+{m_0}^2}$ and $E_y=\pm\sqrt{(\epsilon k_y)^2+{m_0}^2}$ combine to produce $E=\pm\sqrt{(\epsilon k_x)^2+(\epsilon k_y)^2+2{m_0}^2}$.
Here the state propagates both in the $x$ and $y$ directions.
\end{enumerate}
We thereby find a zero-energy eigenvalue, the energy continua starting from $\pm m_0$ and the energy continua starting from $\pm 2m_0$;
see Fig.~\ref{figA1}(c).

We can find the zero-energy eigenfunction based on Eq.~\eqref{zeroeig}.
The zero-energy eigenfunction of $H_{\dirac_x}$ with Eq.~\eqref{wallxB} takes the form of the bound state~\eqref{edge1D} in the $x$ direction and the one of $H_{\dirac_y}$ with Eq.~\eqref{wallyB} takes the similar form of bound state in the $y$ direction.
Multiplying them, we obtain the zero-energy eigenfunction of the 2D Hamiltonian in the form
\begin{align}
\psi_{E=0}^{\dirac, (2)}(x,y)%
&:=\ip{x}{\psi_{E_x=0}^{\dirac, (1)}}\ip{y}{\psi_{E_y=0}^{\dirac, (1)}} \notag \\
&=\frac{m_0}{2\epsilon}\mqty(
1 \\
1
)\otimes \mqty(
1 \\
1
)
\exp\qty[-\frac{m_0}{\epsilon}\qty(|x|+|y|)].
\end{align}
This has a peak in the $xy$ plane because an edge state running in the $x$ direction and the edge state running in the $y$ direction were multiplied together, and hence is identified as a second-order topological state, namely the corner state.
This describes each peak that our 2D DTQW demonstrates in Fig.~\ref{corner}.

\section{\label{appendmomotaro}Band Structure for Finite $\theta_x$ and $\theta_y$}

We here show the band structure for finite $\theta_x$ and $\theta_y$.
In the following, let us fix $\theta_x=\pi/3$ and vary $\theta_y$ from 0 to $\pi/3$.

\begin{figure*}[t]
\begin{minipage}[b]{0.48\textwidth}
\vspace{0mm}
\centering
\includegraphics[width=\textwidth]{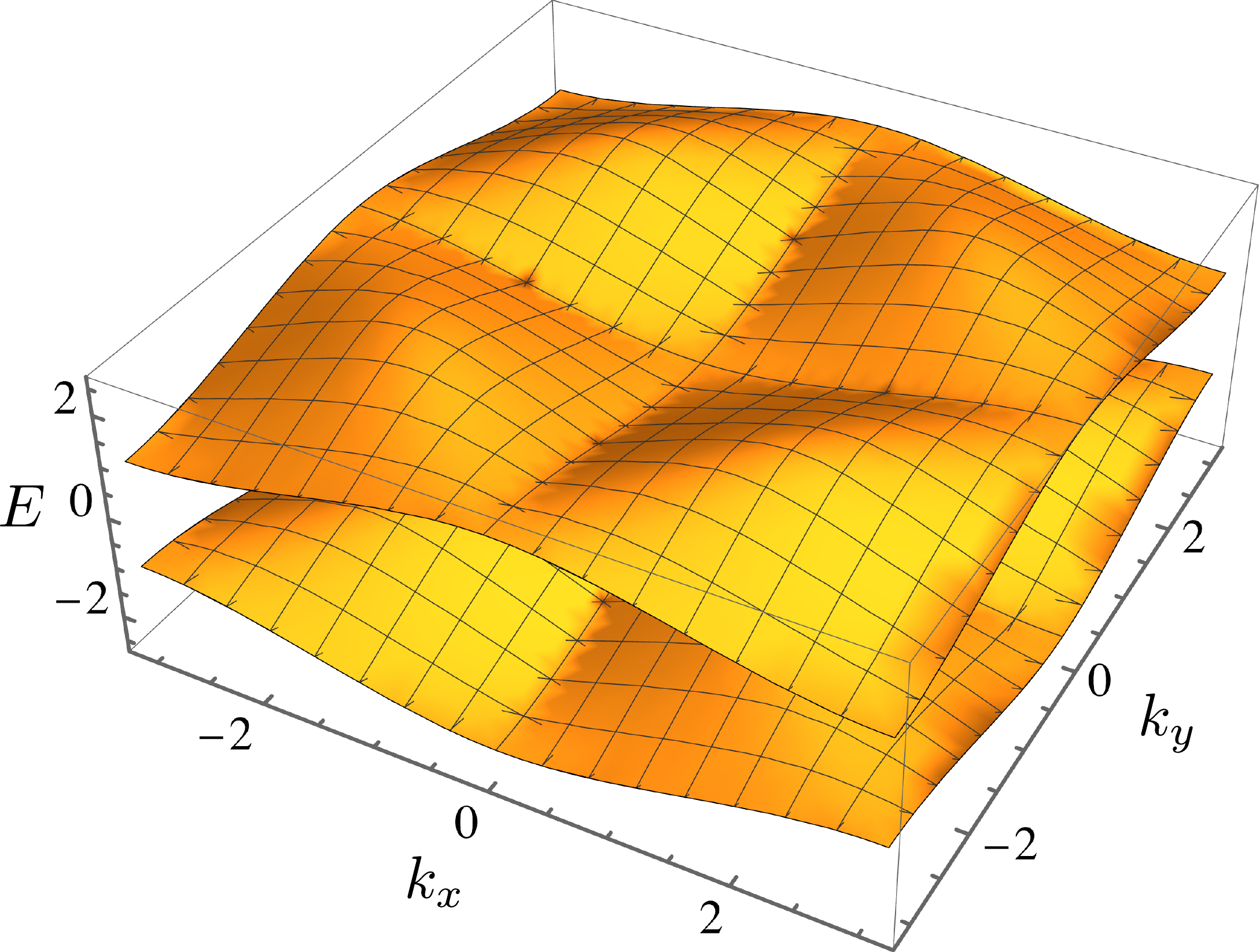}
\\
(a)
\end{minipage}
\hfill
\begin{minipage}[b]{0.48\textwidth}
\centering
\includegraphics[width=\textwidth]{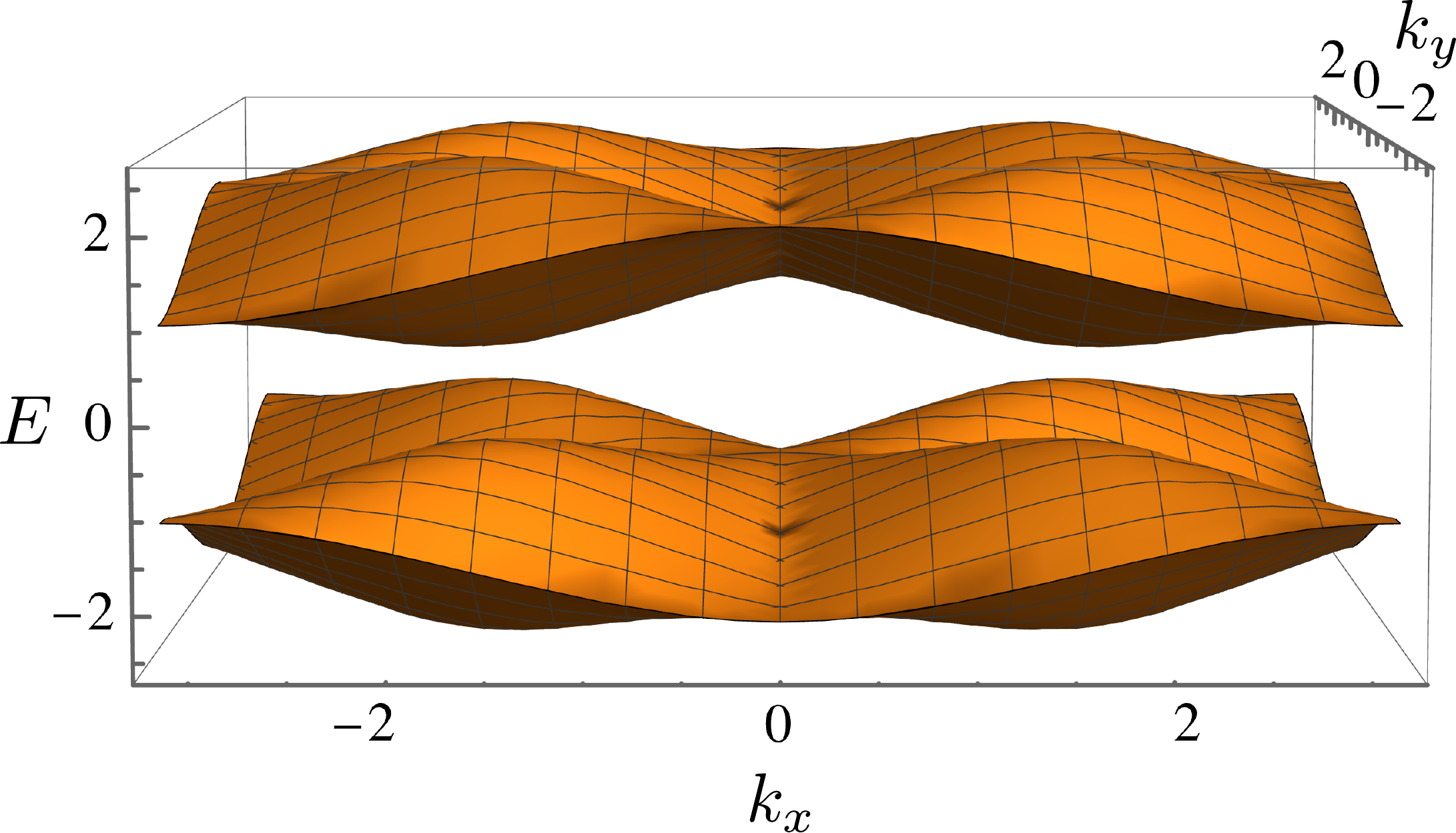}
\\
(b)
\end{minipage}
\\[2\baselineskip]
\begin{minipage}[b]{0.48\textwidth}
\vspace{0mm}
\centering
\includegraphics[width=\textwidth]{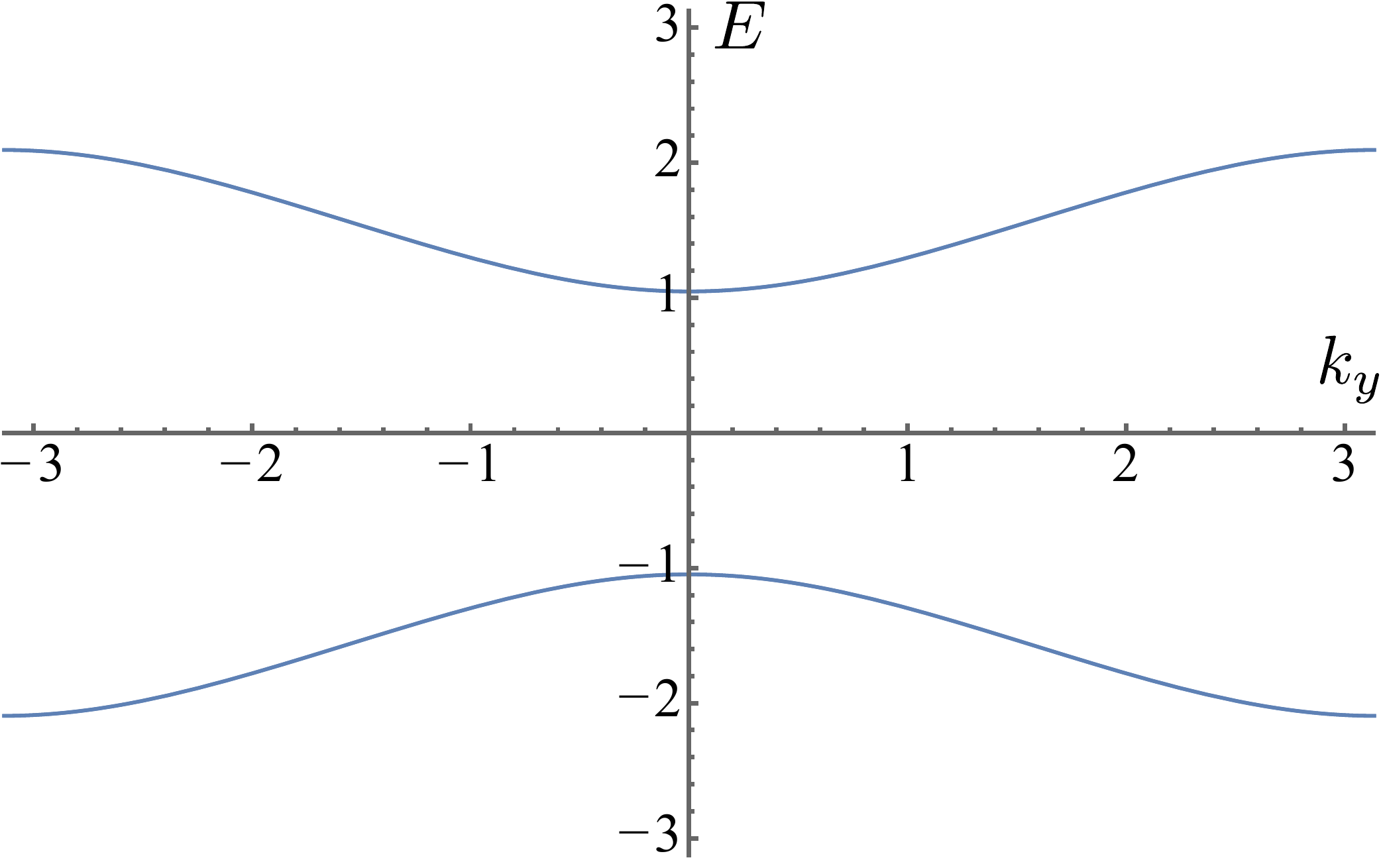}
\\
(c)
\end{minipage}
\hfill
\begin{minipage}[b]{0.48\textwidth}
\centering
\includegraphics[width=\textwidth]{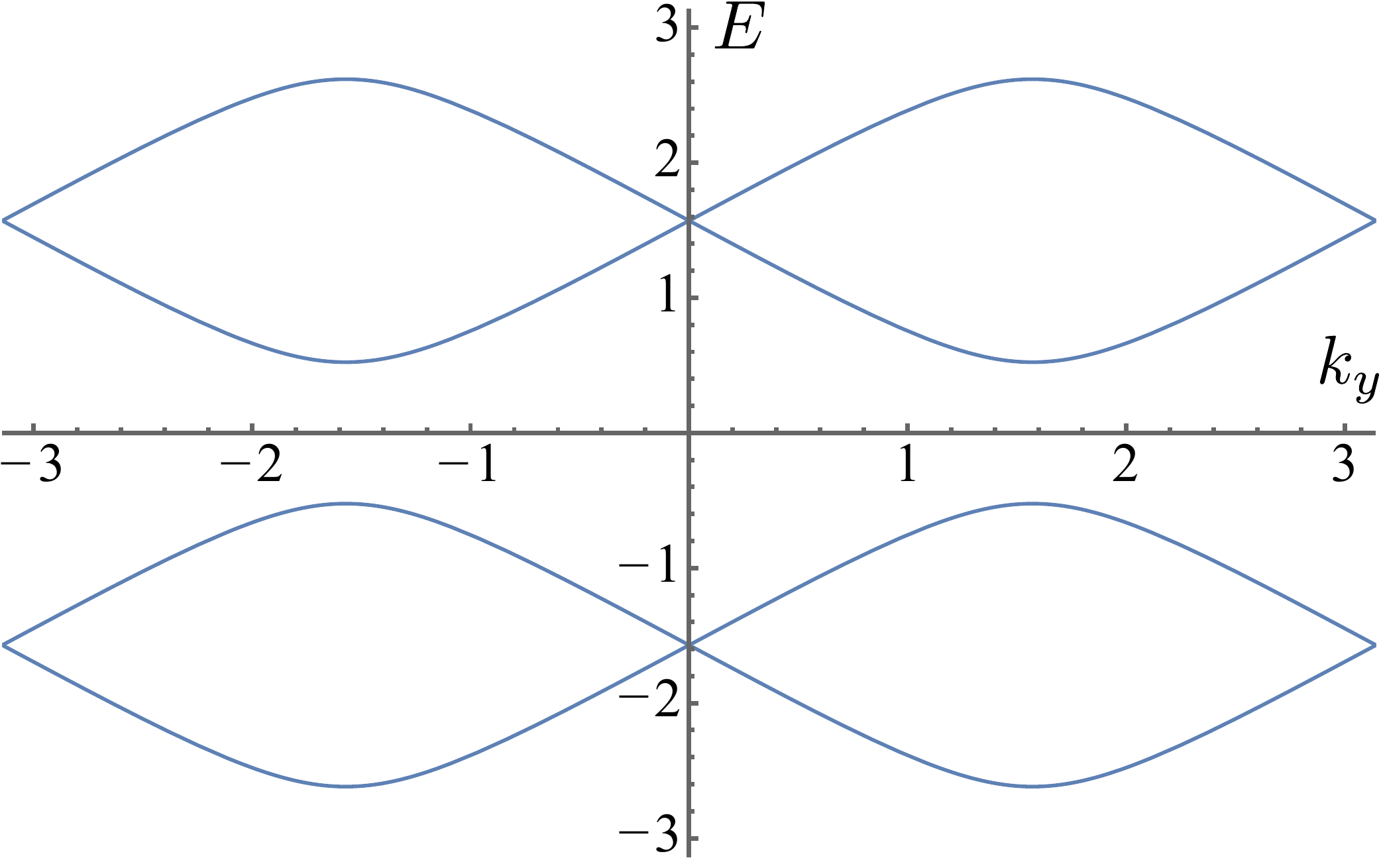}
\\
(d)
\end{minipage}
\\[2\baselineskip]
\begin{minipage}[b]{0.48\textwidth}
\vspace{0mm}
\centering
\includegraphics[width=\textwidth]{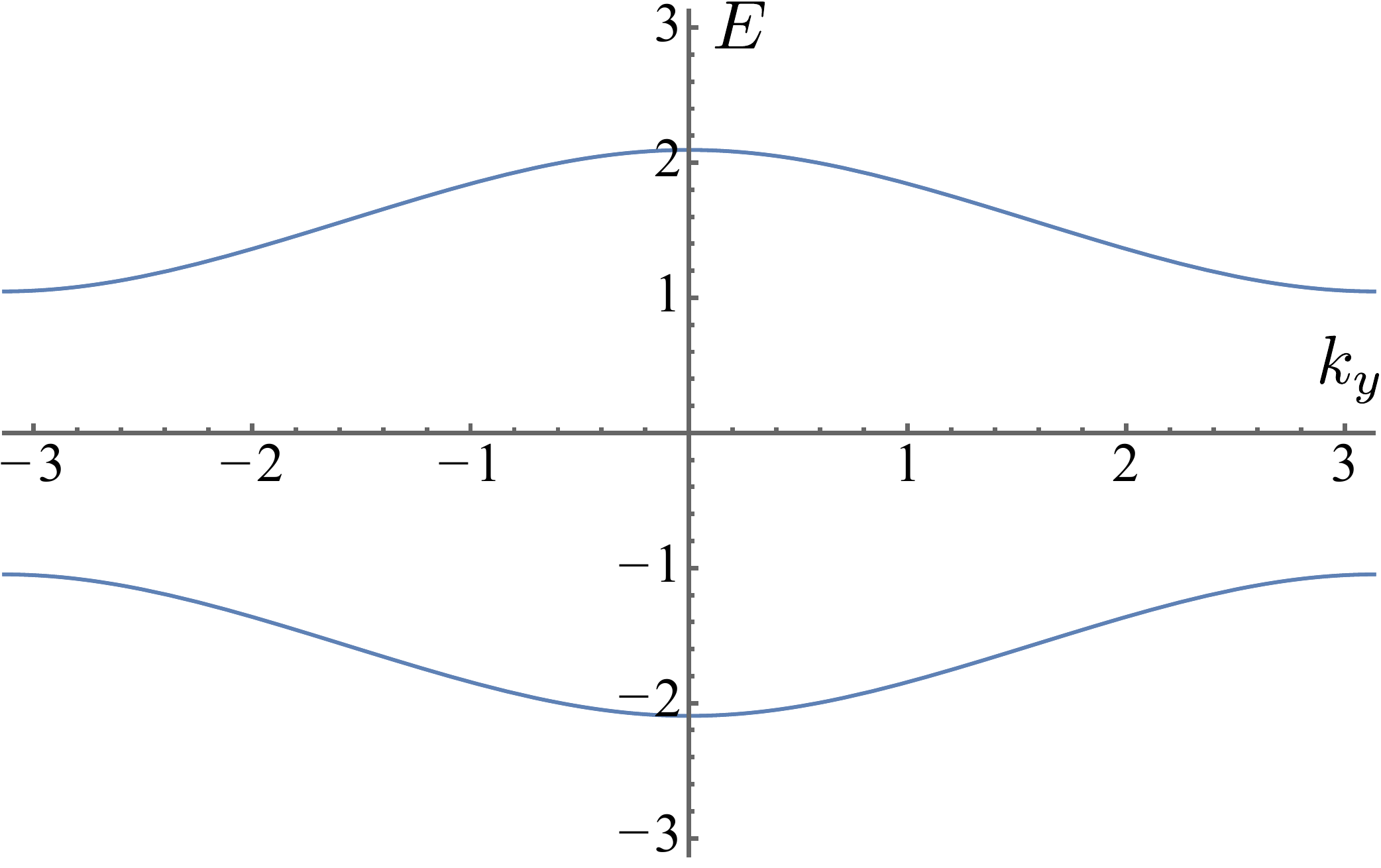}
\\
(e)
\end{minipage}
\hfill
\begin{minipage}[b]{0.48\textwidth}
\centering
\includegraphics[width=\textwidth]{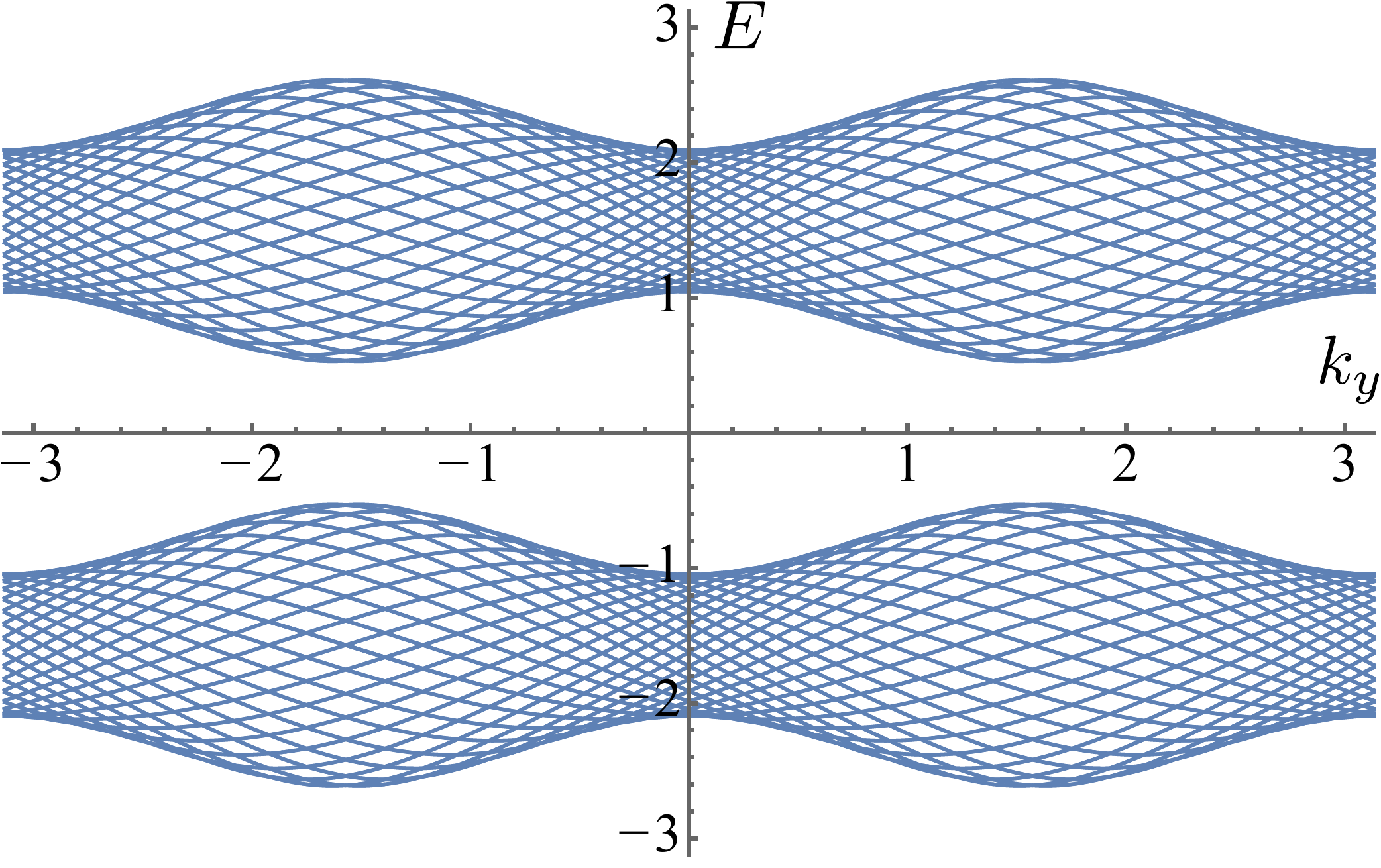}
\\
(f)
\end{minipage}
\caption{Band structure for $\theta_x=\pi/3$ and $\theta_y=0$.
(a) and (b) Energy bands from two different viewpoints.
(c), (d) and (e) Cross sections of the bands at $k_x=0$, $k_x=\pi/2$ and $k_x=\pi$, respectively.
(f) Projection of the bands over the $k_x$ axis onto the $k_y$ axis. We set $\hbar=a=\Delta t=1$.}
\label{figS1}
\end{figure*}

The model generally has four bands.
For $\theta_y=0$, the first and second bands as well as the third and fourth bands are closed on the lines $k_x=0,\pm\pi$ and $k_y=0,\pm\pi$;
see Fig.~\ref{figS1}.
When projected on the $k_y$ axis, each of the upper and lower energy band appears to be filled.

\begin{figure*}[t]
\begin{minipage}[b]{0.48\textwidth}
\vspace{0mm}
\centering
\includegraphics[width=\textwidth]{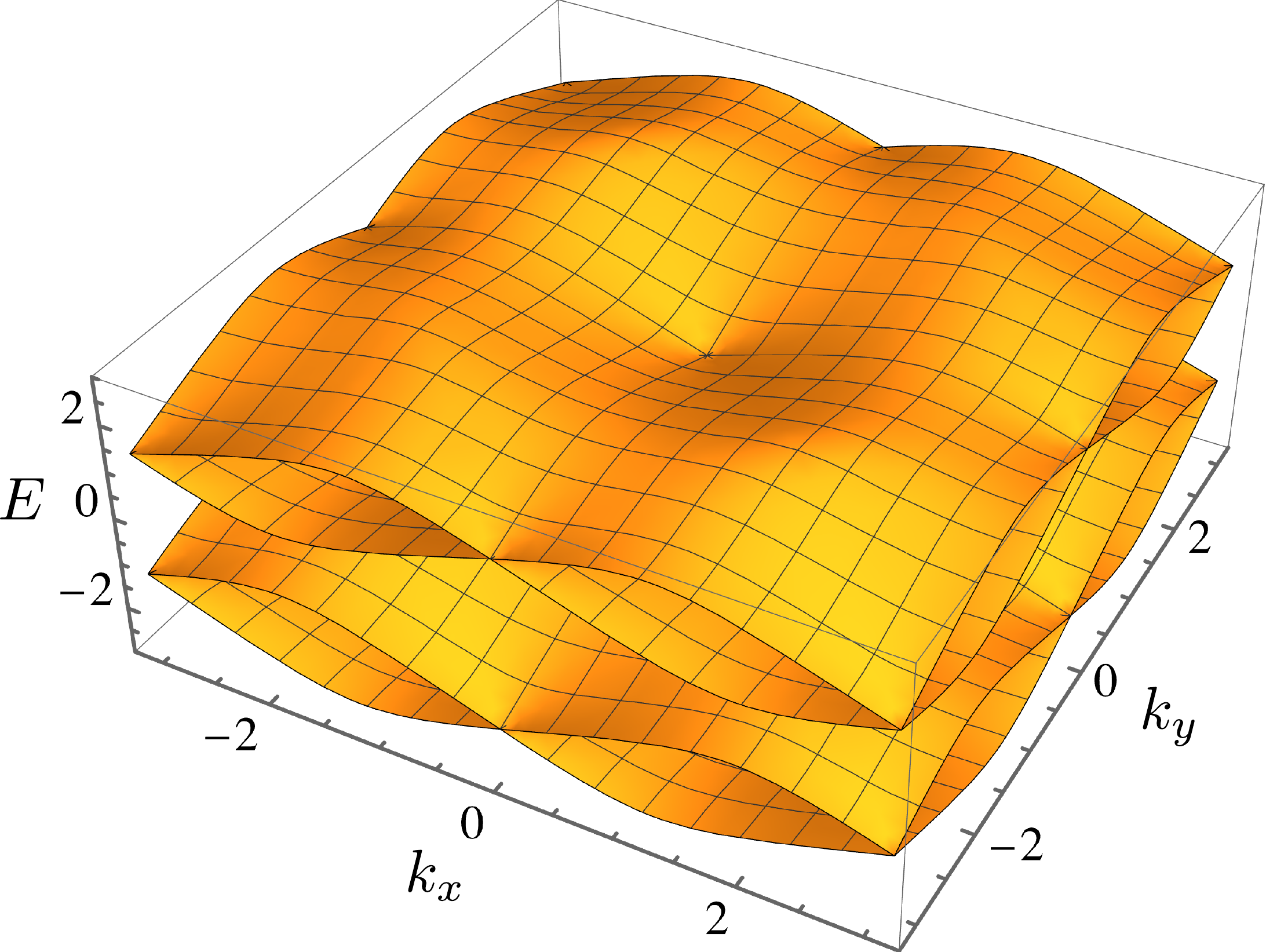}
\\
(a)
\end{minipage}
\hfill
\begin{minipage}[b]{0.48\textwidth}
\centering
\includegraphics[width=\textwidth]{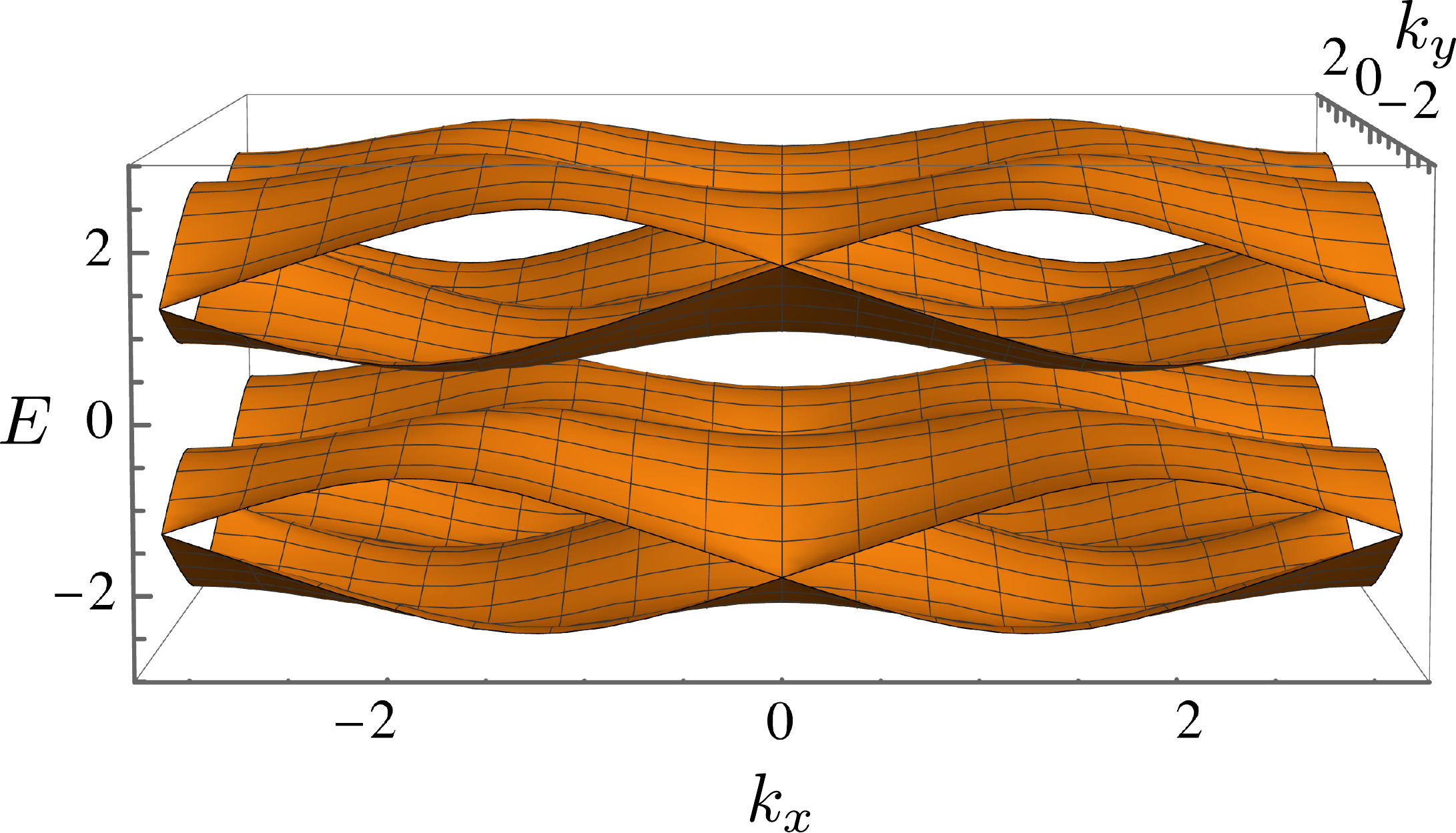}
\\
(b)
\end{minipage}
\\[2\baselineskip]
\begin{minipage}[b]{0.48\textwidth}
\vspace{0mm}
\centering
\includegraphics[width=\textwidth]{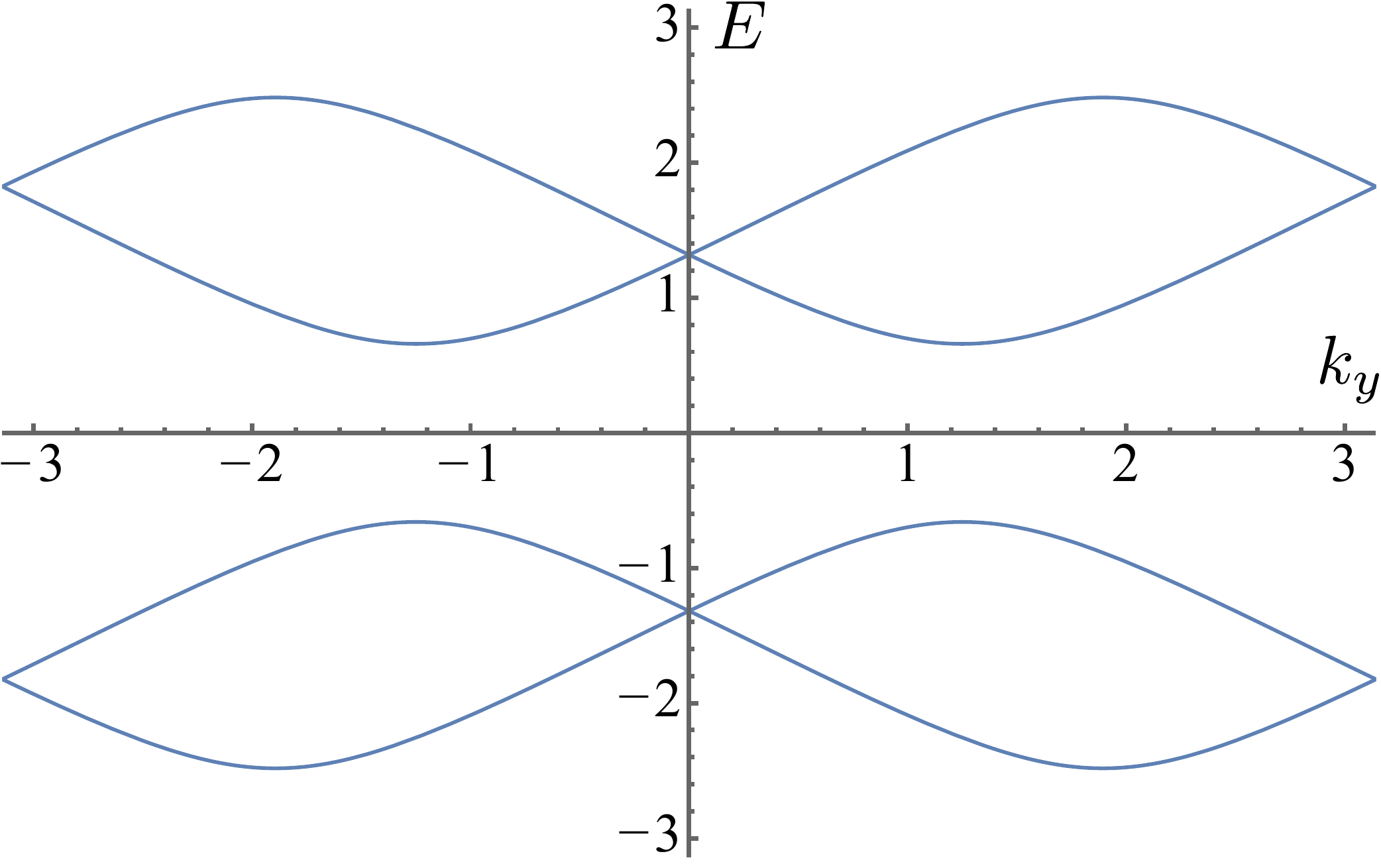}
\\
(c)
\end{minipage}
\hfill
\begin{minipage}[b]{0.48\textwidth}
\centering
\includegraphics[width=\textwidth]{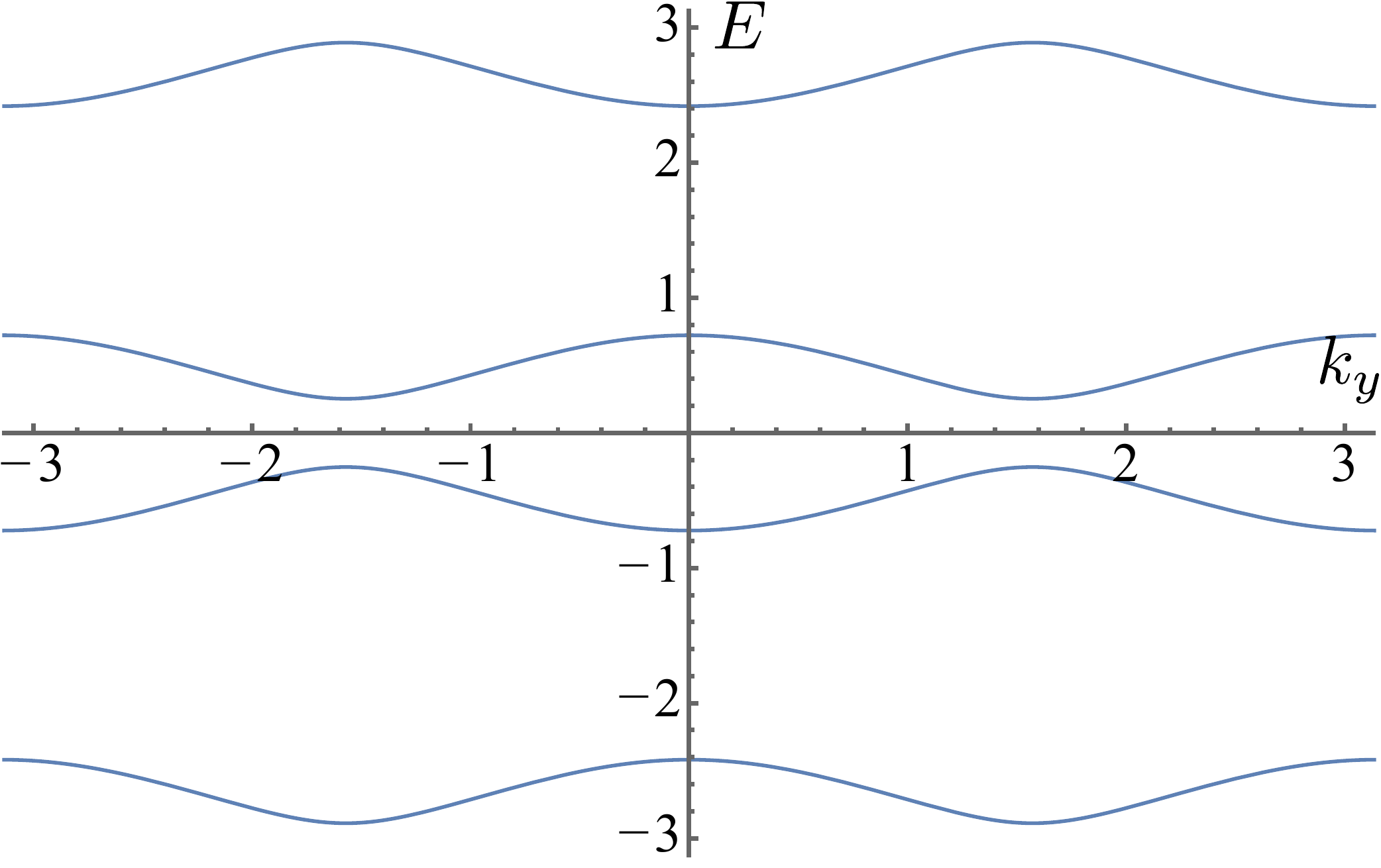}
\\
(d)
\end{minipage}
\\[2\baselineskip]
\begin{minipage}[b]{0.48\textwidth}
\vspace{0mm}
\centering
\includegraphics[width=\textwidth]{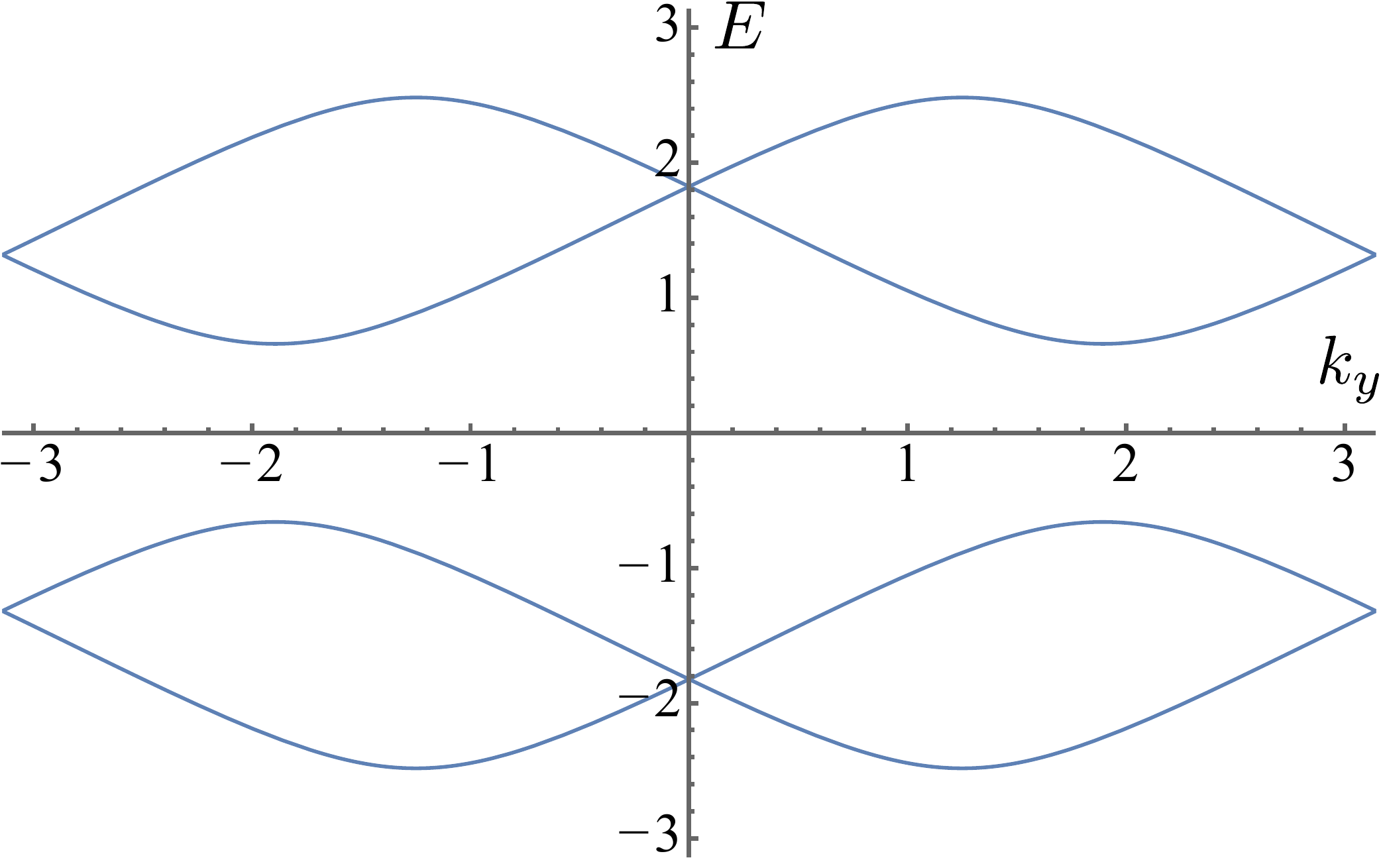}
\\
(e)
\end{minipage}
\hfill
\begin{minipage}[b]{0.48\textwidth}
\centering
\includegraphics[width=\textwidth]{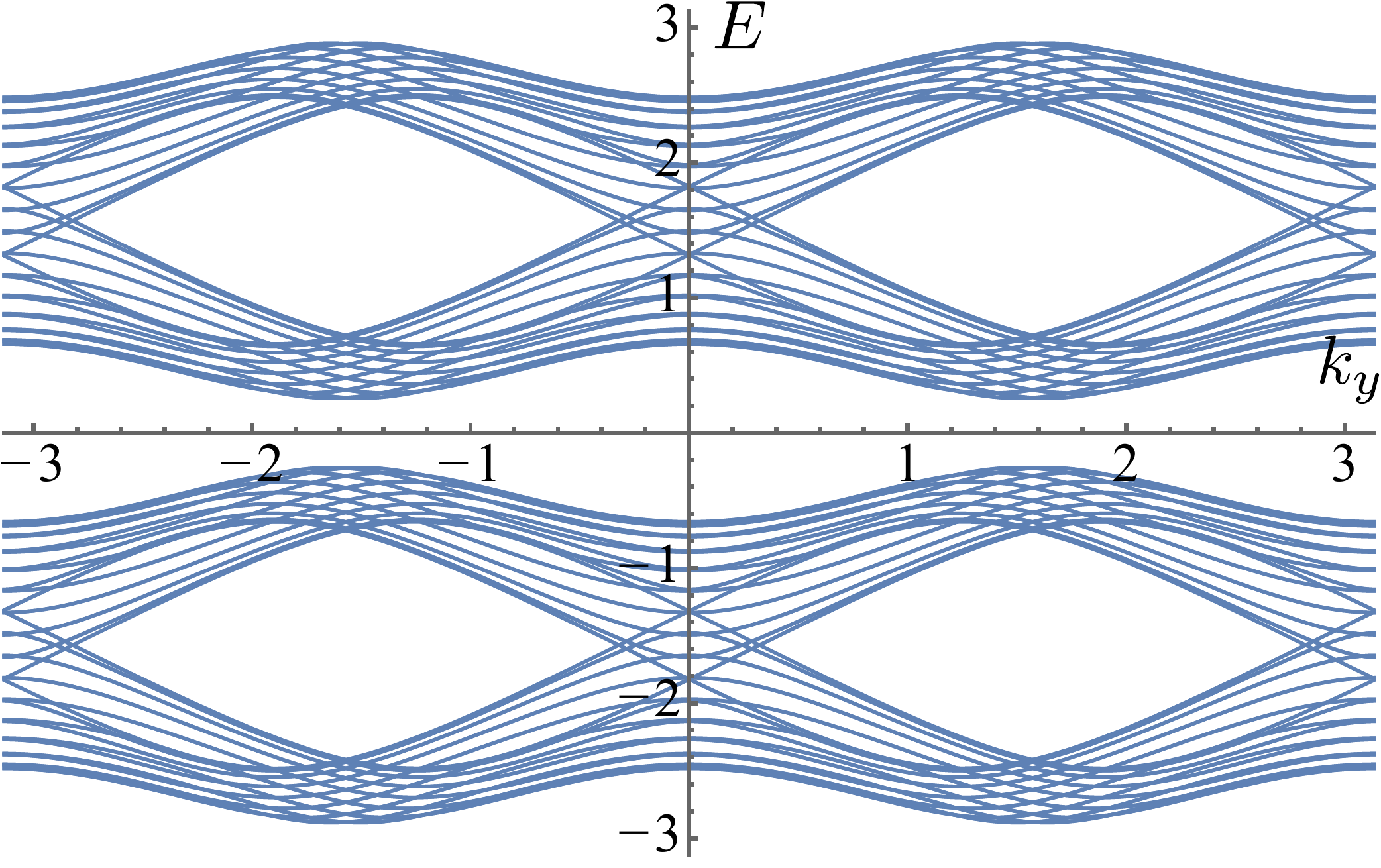}
\\
(f)
\end{minipage}
\caption{Band structure for $\theta_x=\pi/3$ and $\theta_y=\pi/3$.
(a) and (b) Energy bands from two different viewpoints.
(c), (d) and (e) Cross sections of the bands at $k_x=0$, $k_x=\pi/2$ and $k_x=\pi$, respectively.
(f) Projection of the bands over the $k_x$ axis onto the $k_y$ axis. We set $\hbar=a=\Delta t=1$.}
\label{figS2}
\end{figure*}

For $\theta_y=\pi/3$, on the other hand, all bands are open on the lines $k_x=0,\pm\pi$ and $k_y=0,\pm\pi$ except for the Dirac points $(k_x,k_y)=(0,0),(\pm \pi/2,\pm \pi/2), (\pm \pi,\pm,\pi)$;
see Fig.~\ref{figS2}.
When projected on the $k_y$ axis, we can now see large openings in each of the upper and lower bands.
When we introduce the effective potential
\begin{align}
\theta_x(x)=\begin{cases}
-\pi/3 & \quad\mbox{for $|x|>L_x/2$},
\\
\pi/3 & \quad\mbox{for $|x|\le L_x/2$},
\end{cases}
\end{align}
we see edge modes in the openings as in Fig.~7 of the main text.

\begin{figure*}
\begin{minipage}[t]{0.32\textwidth}
\centering
\includegraphics[width=\textwidth]{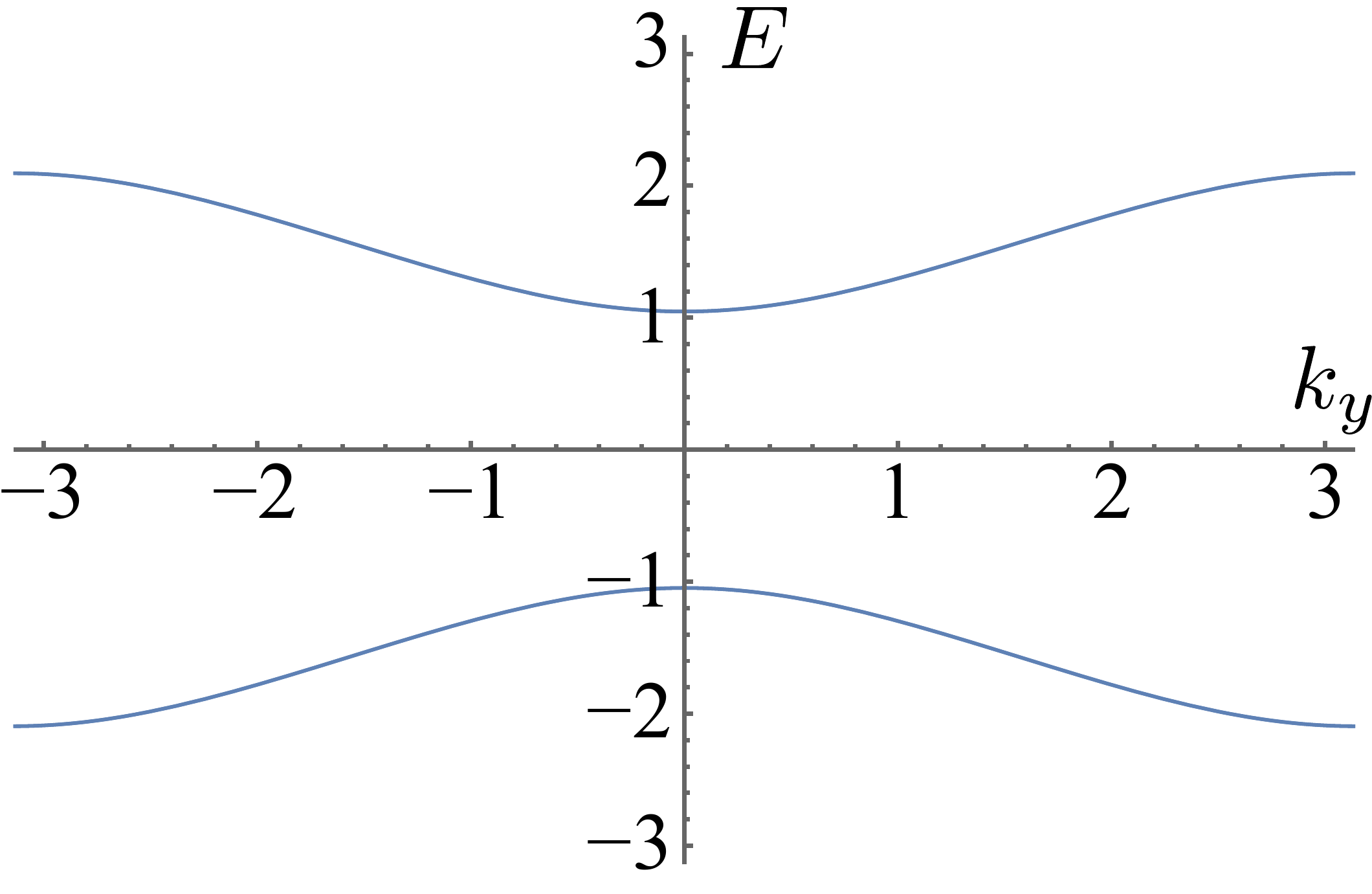}
\\
(a)
\end{minipage}
\hfill
\begin{minipage}[t]{0.32\textwidth}
\centering
\includegraphics[width=\textwidth]{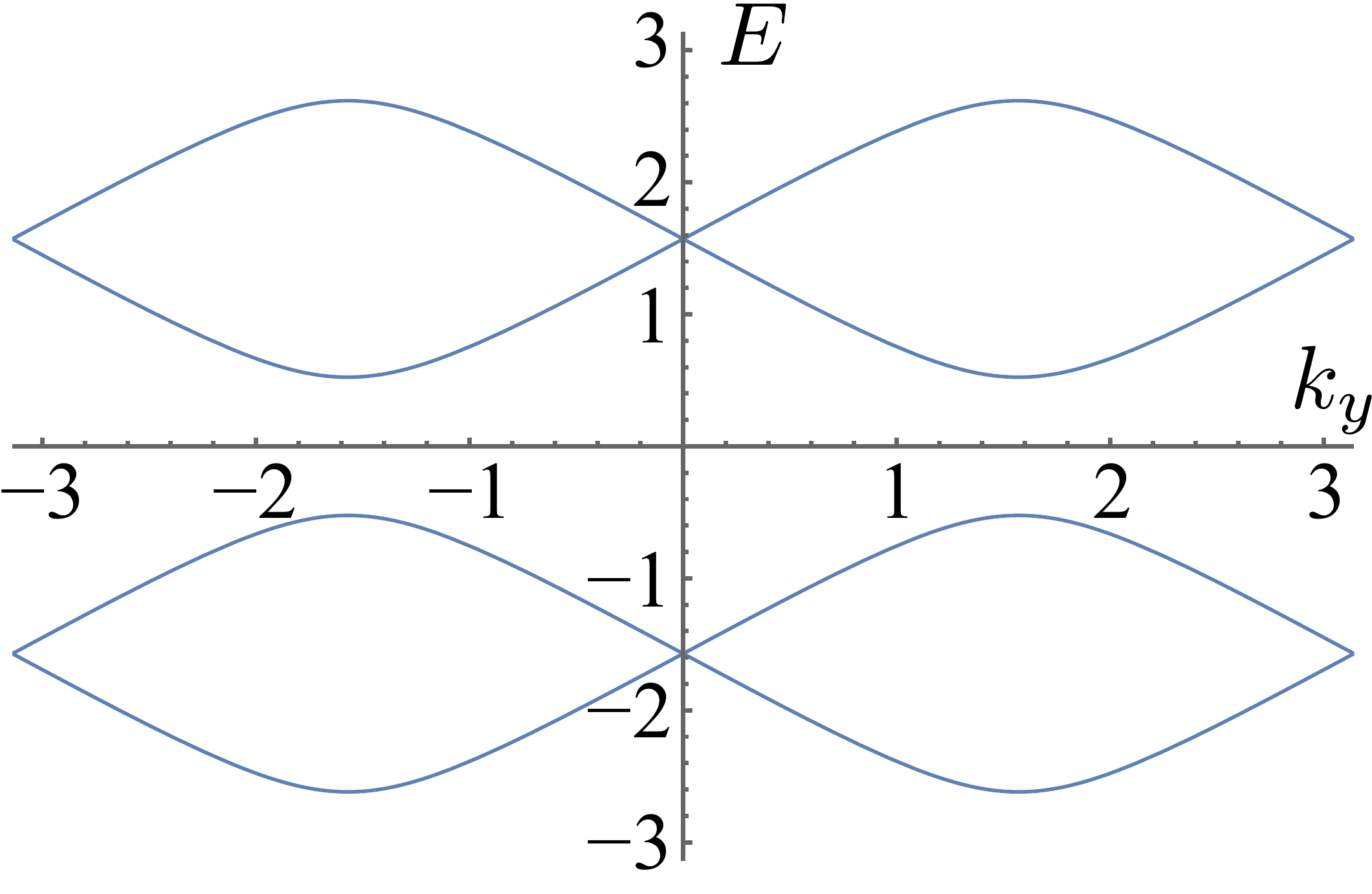}
\\
(b)
\end{minipage}
\hfill
\begin{minipage}[t]{0.32\textwidth}
\centering
\includegraphics[width=\textwidth]{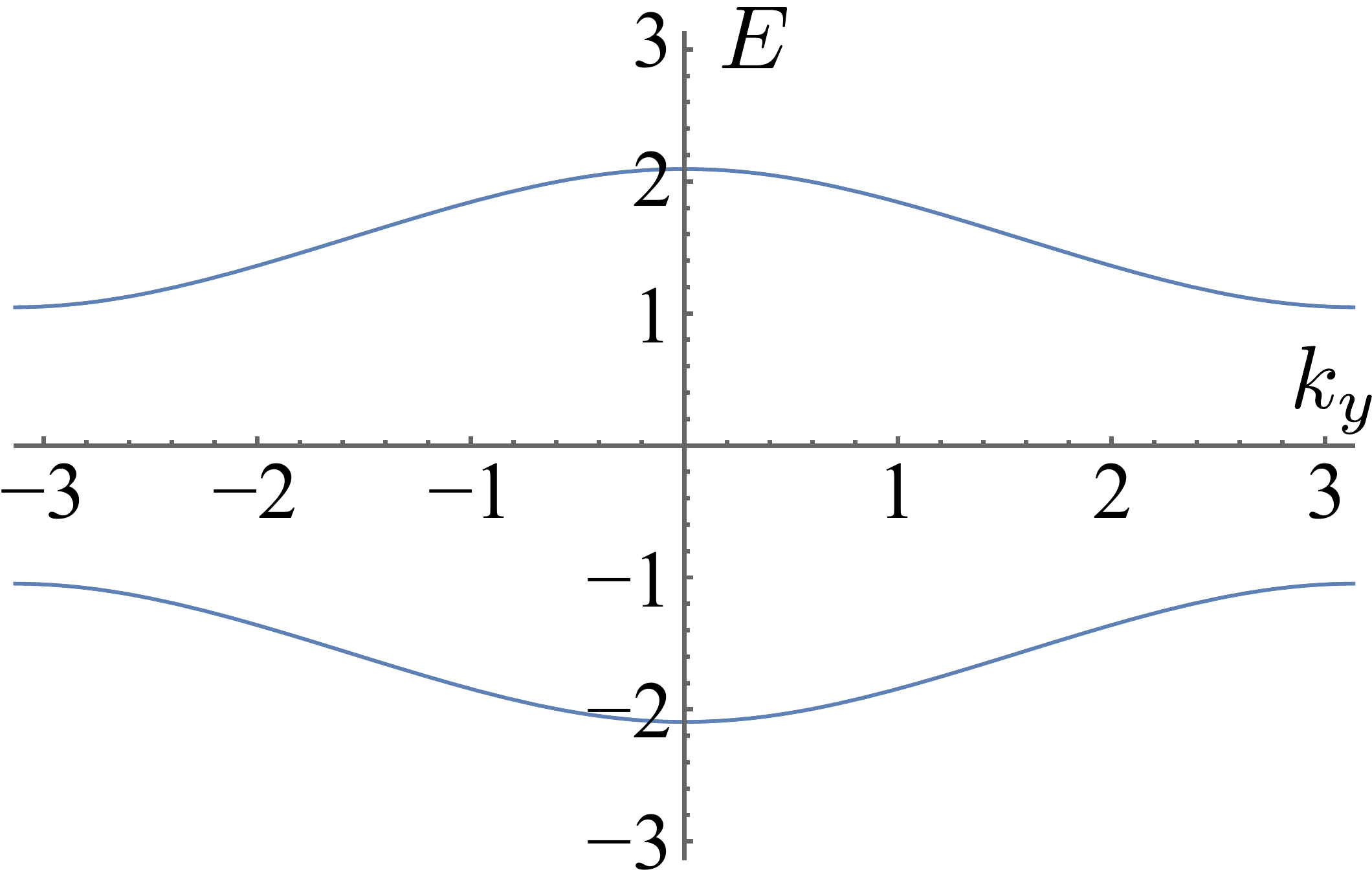}
\\
(c)
\end{minipage}
\\[\baselineskip]
\begin{minipage}[t]{0.32\textwidth}
\centering
\includegraphics[width=\textwidth]{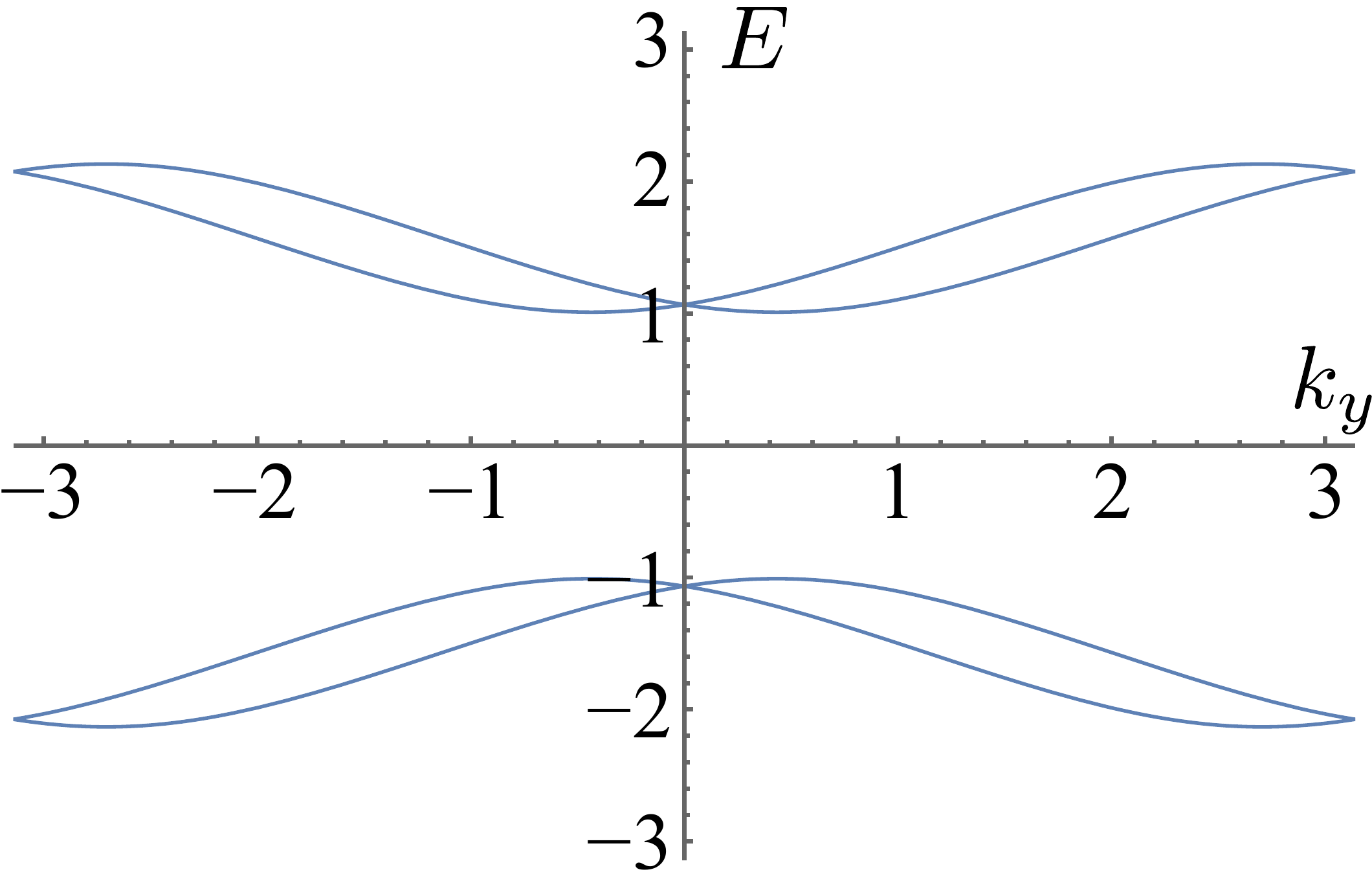}
\\
(d)
\end{minipage}
\hfill
\begin{minipage}[t]{0.32\textwidth}
\centering
\includegraphics[width=\textwidth]{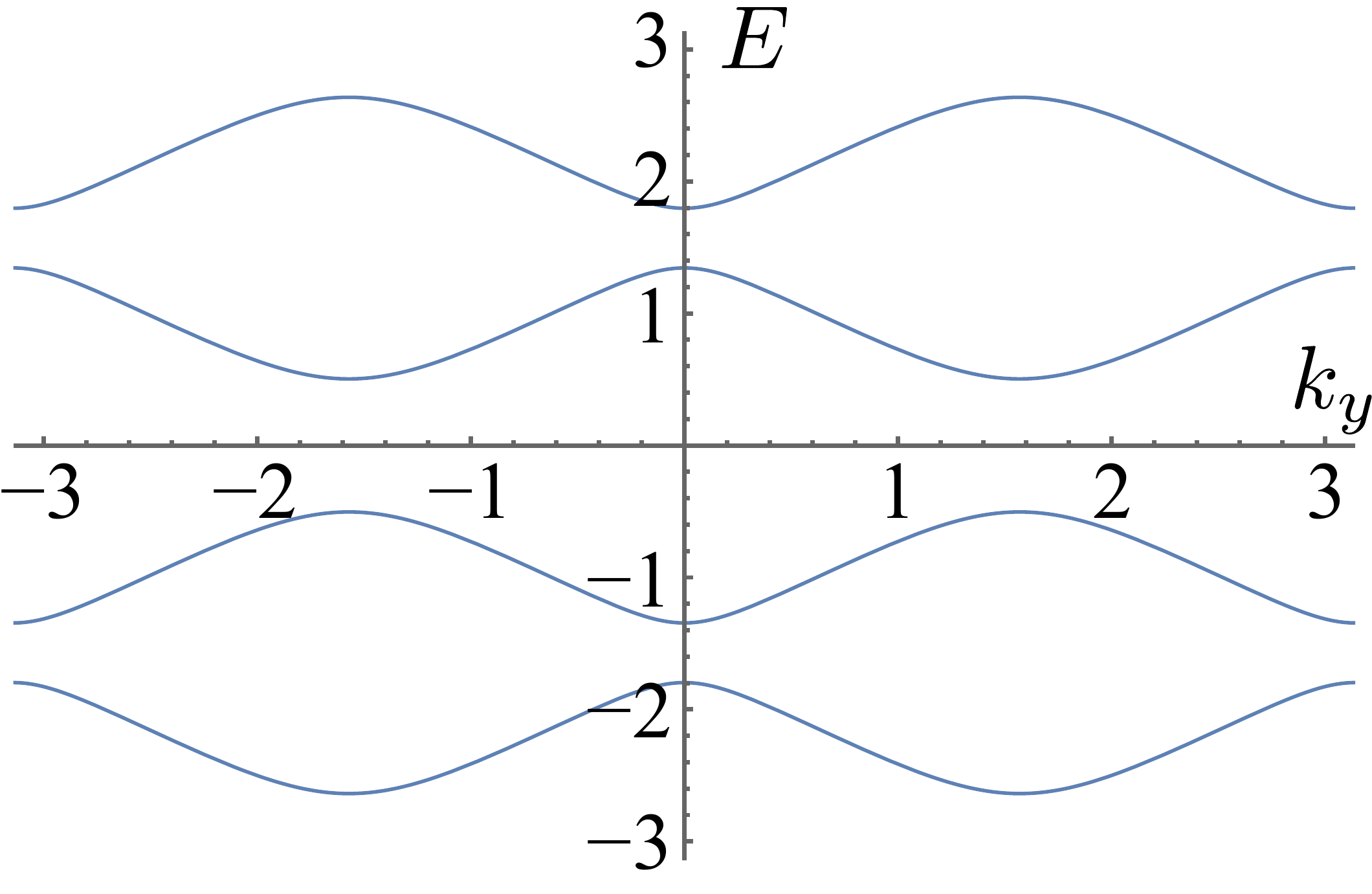}
\\
(e)
\end{minipage}
\hfill
\begin{minipage}[t]{0.32\textwidth}
\centering
\includegraphics[width=\textwidth]{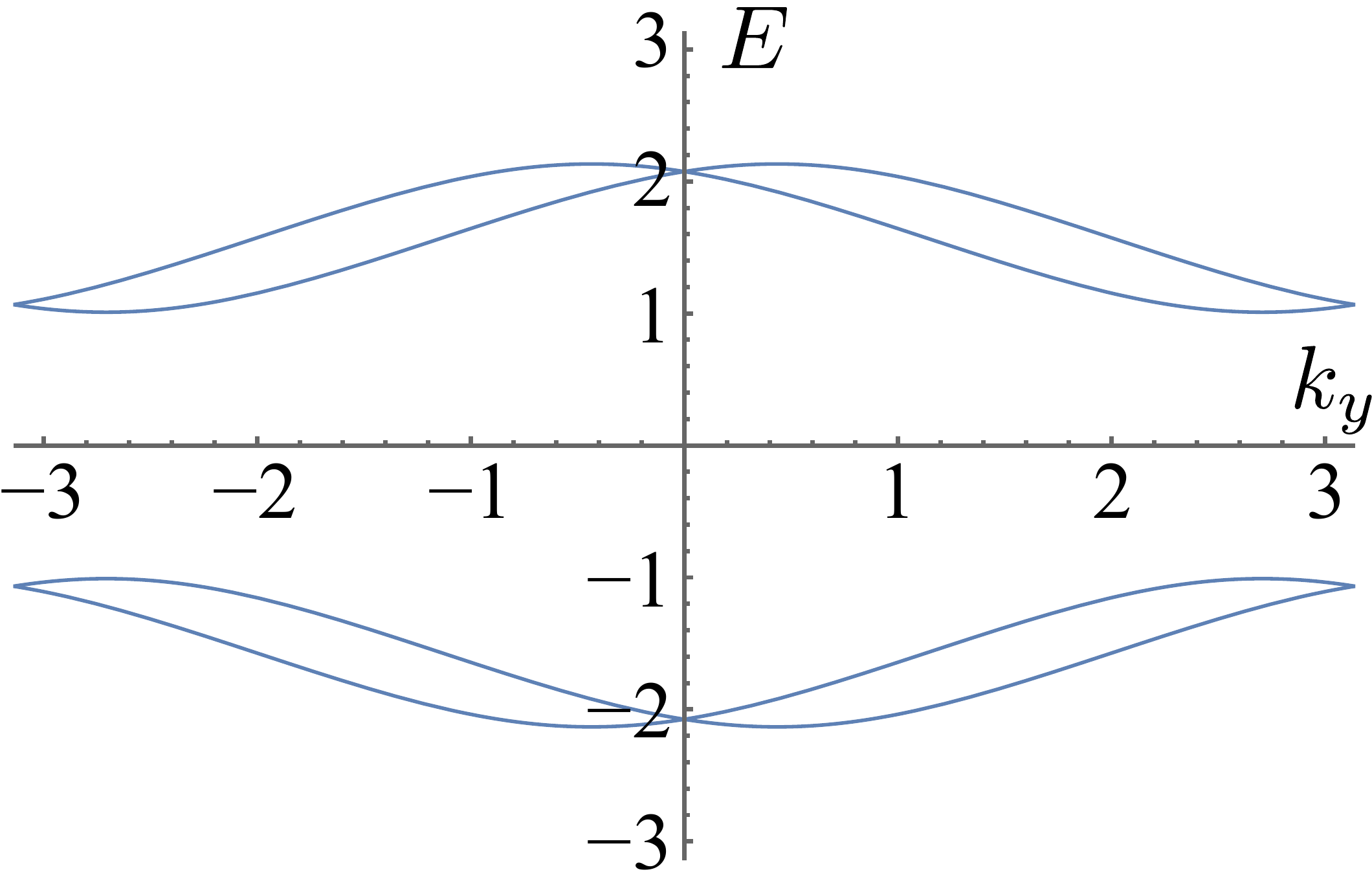}
\\
(f)
\end{minipage}
\\[\baselineskip]
\begin{minipage}[t]{0.32\textwidth}
\centering
\includegraphics[width=\textwidth]{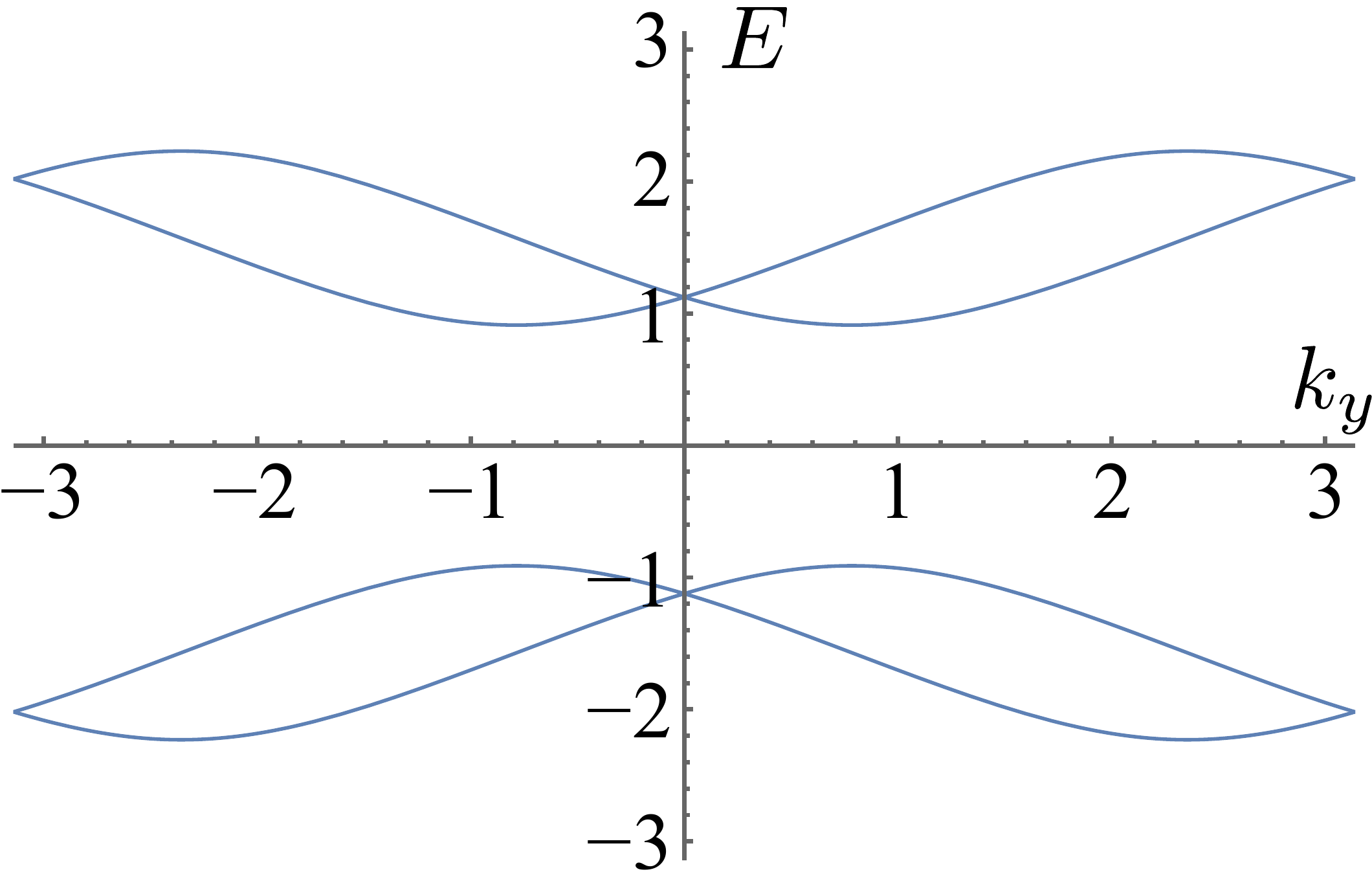}
\\
(g)
\end{minipage}
\hfill
\begin{minipage}[t]{0.32\textwidth}
\centering
\includegraphics[width=\textwidth]{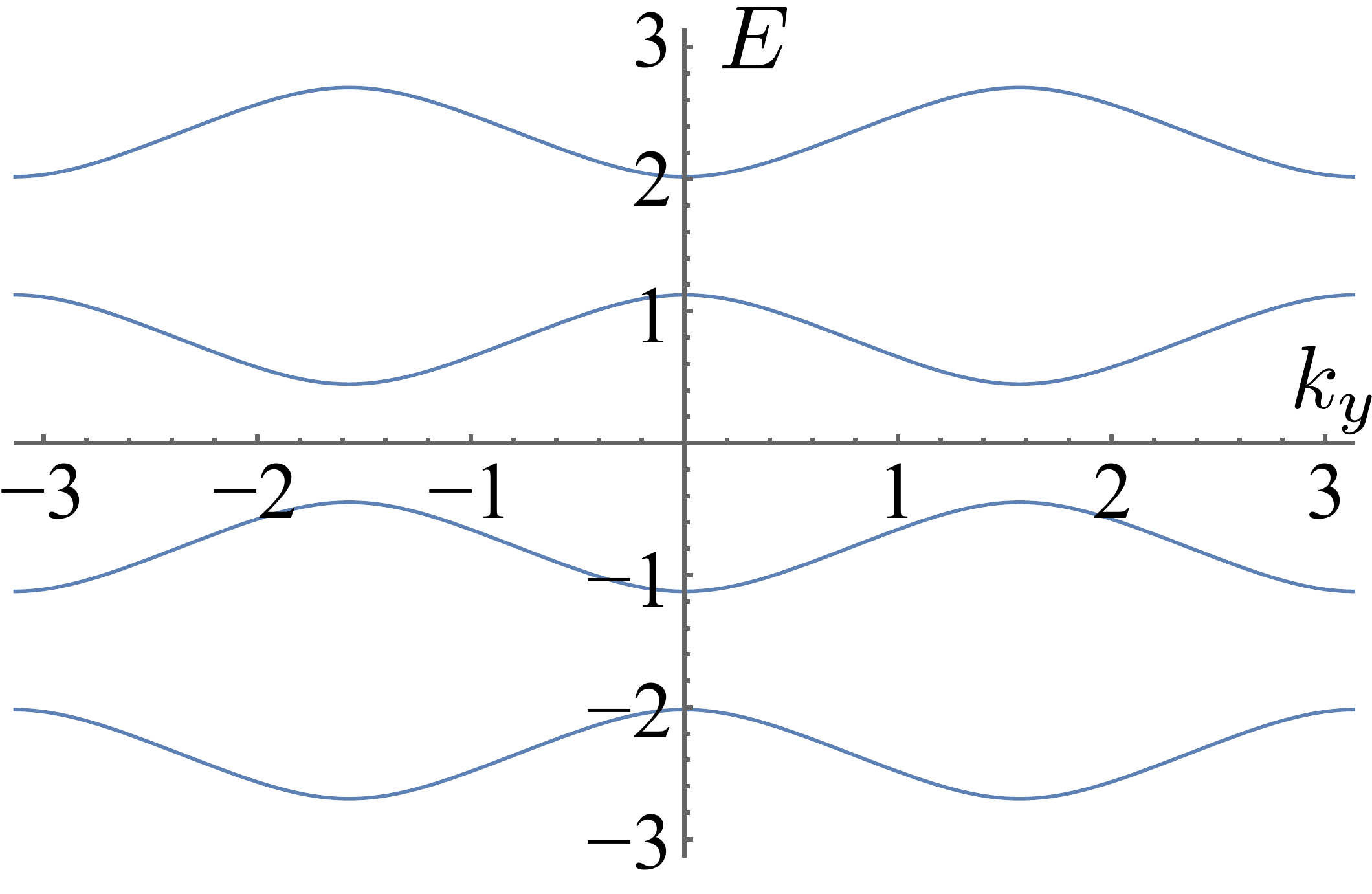}
\\
(h)
\end{minipage}
\hfill
\begin{minipage}[t]{0.32\textwidth}
\centering
\includegraphics[width=\textwidth]{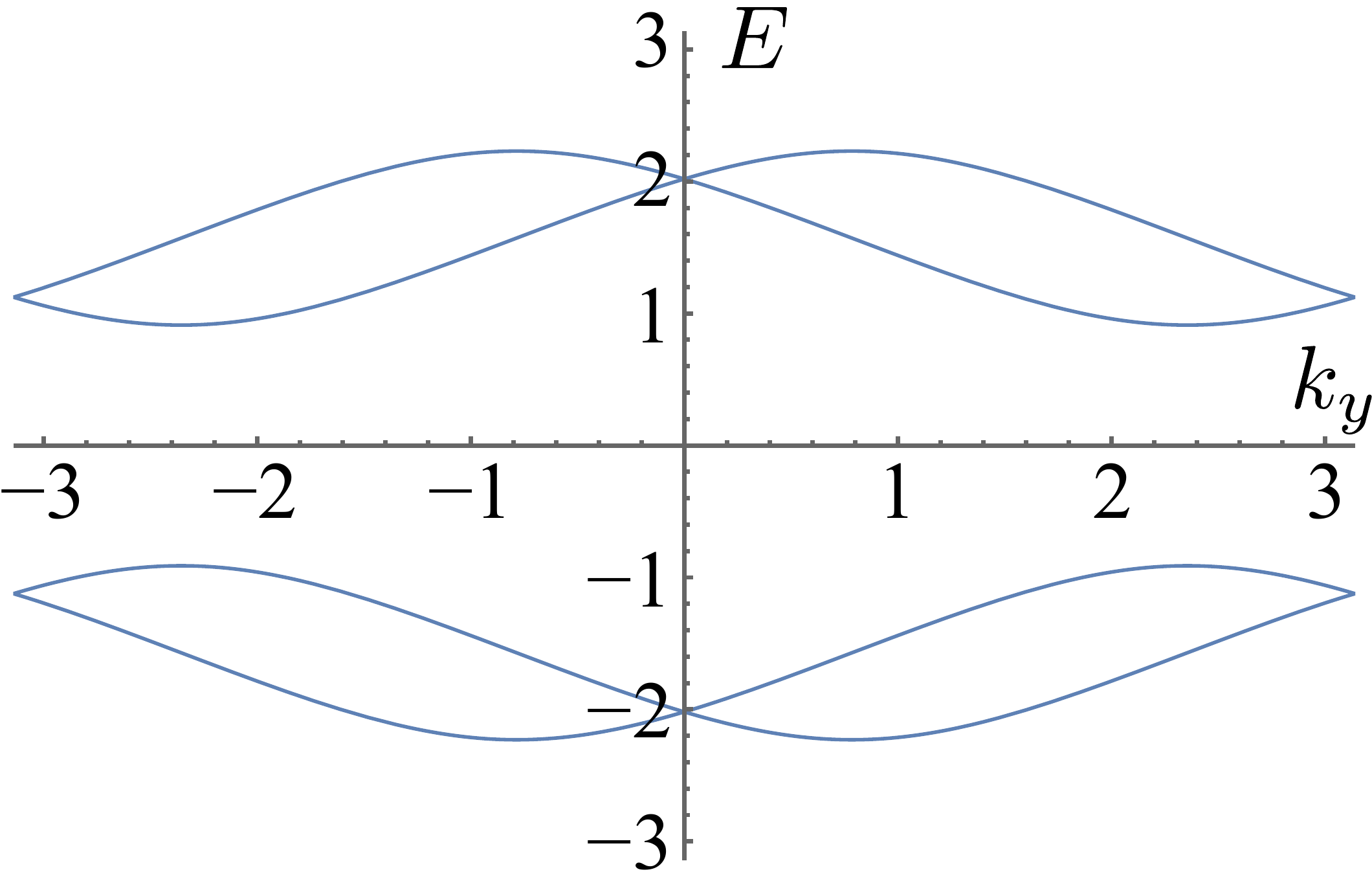}
\\
(i)
\end{minipage}
\\[\baselineskip]
\begin{minipage}[t]{0.32\textwidth}
\centering
\includegraphics[width=\textwidth]{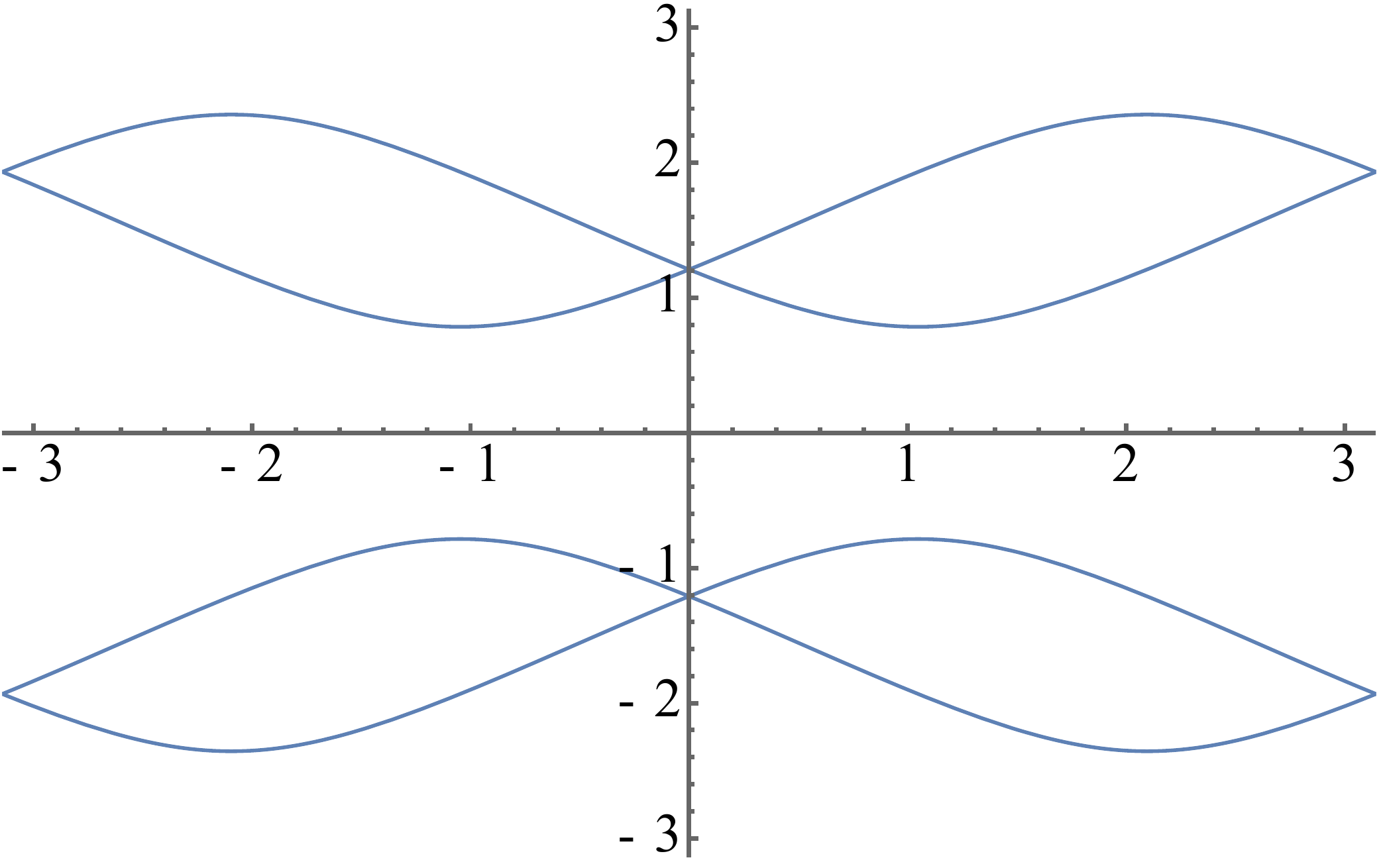}
\\
(j)
\end{minipage}
\hfill
\begin{minipage}[t]{0.32\textwidth}
\centering
\includegraphics[width=\textwidth]{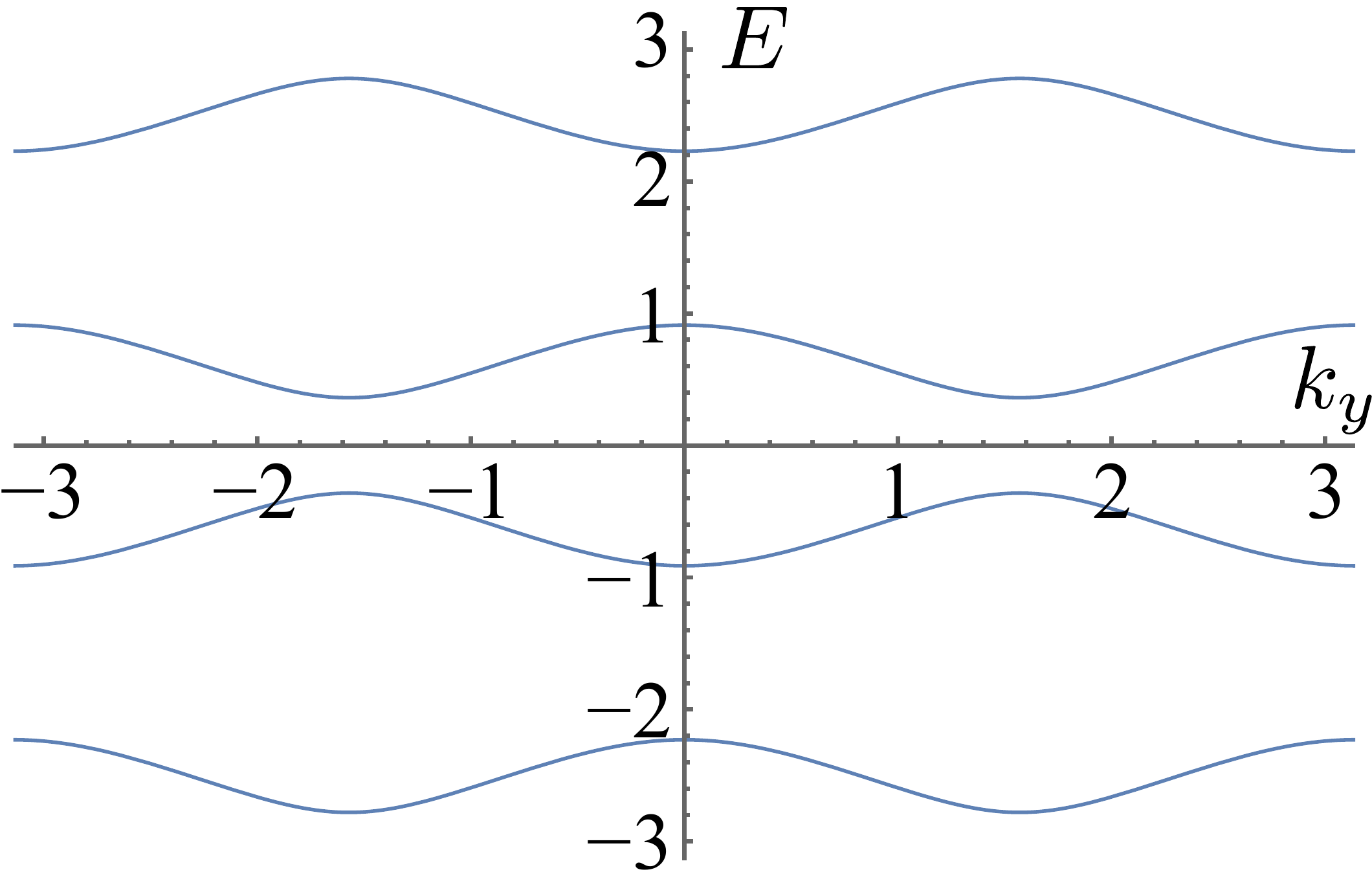}
\\
(k)
\end{minipage}
\hfill
\begin{minipage}[t]{0.32\textwidth}
\centering
\includegraphics[width=\textwidth]{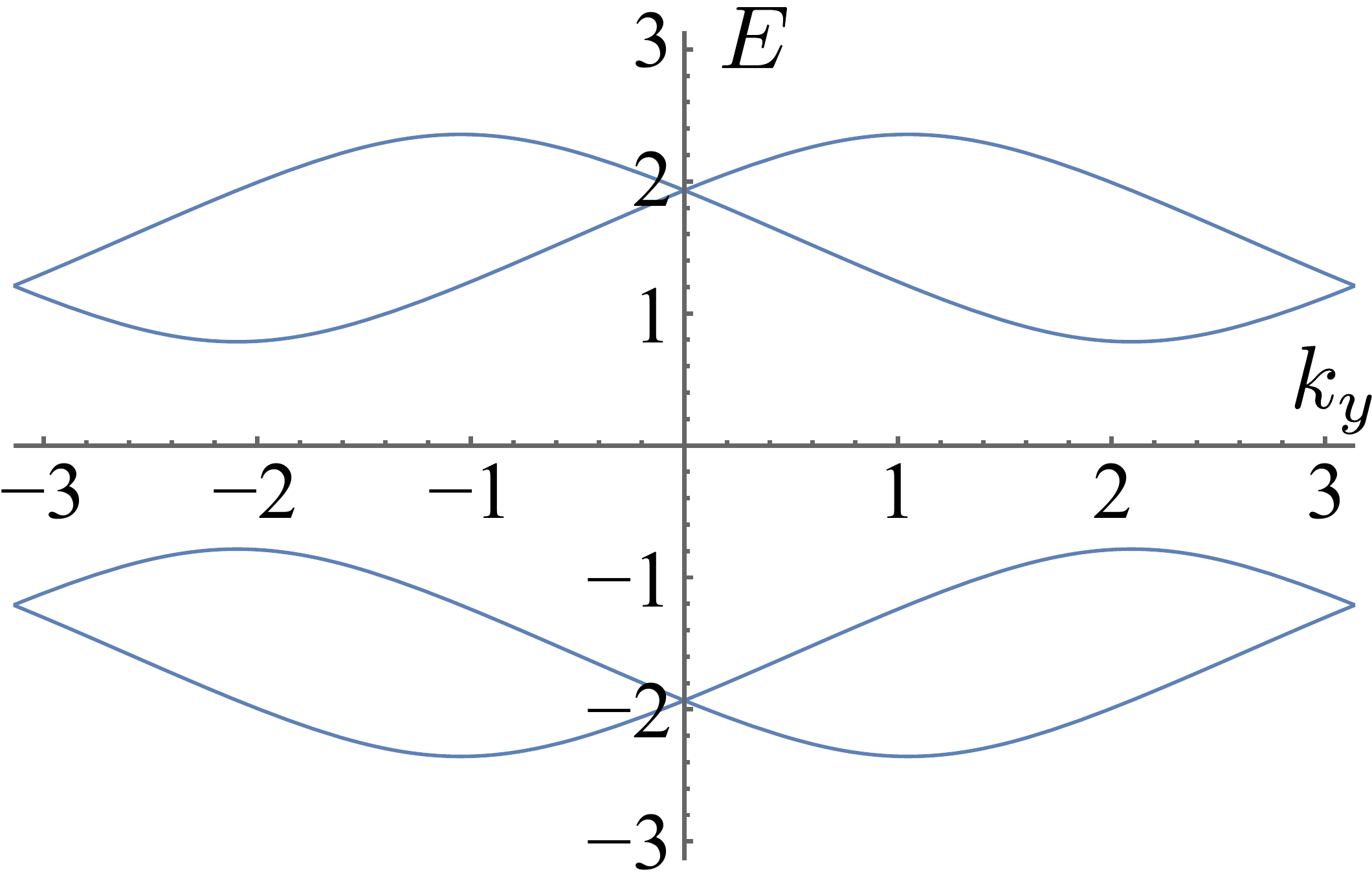}
\\
(l)
\end{minipage}
\\[\baselineskip]
\begin{minipage}[t]{0.32\textwidth}
\centering
\includegraphics[width=\textwidth]{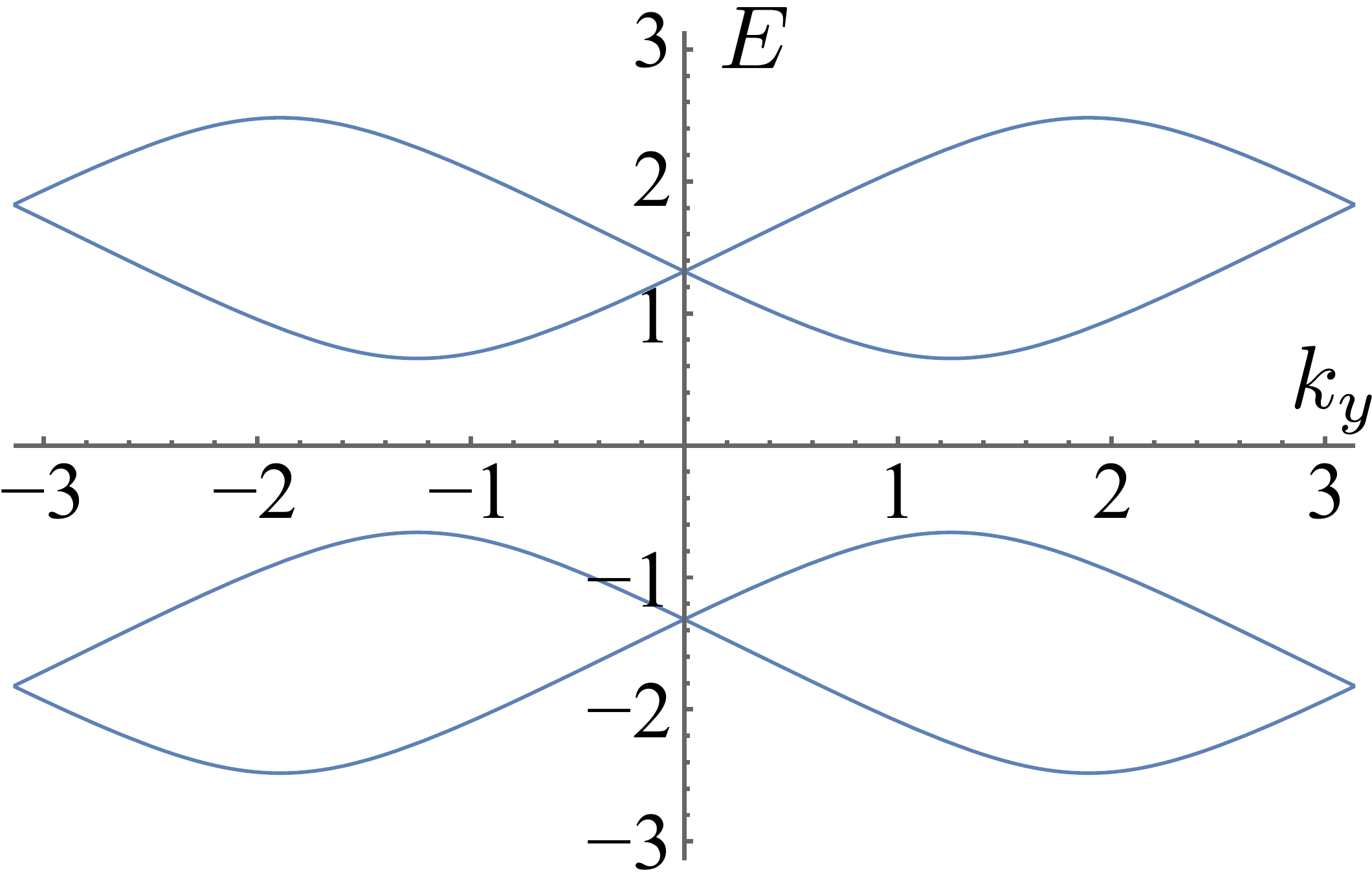}
\\
(m)
\end{minipage}
\hfill
\begin{minipage}[t]{0.32\textwidth}
\centering
\includegraphics[width=\textwidth]{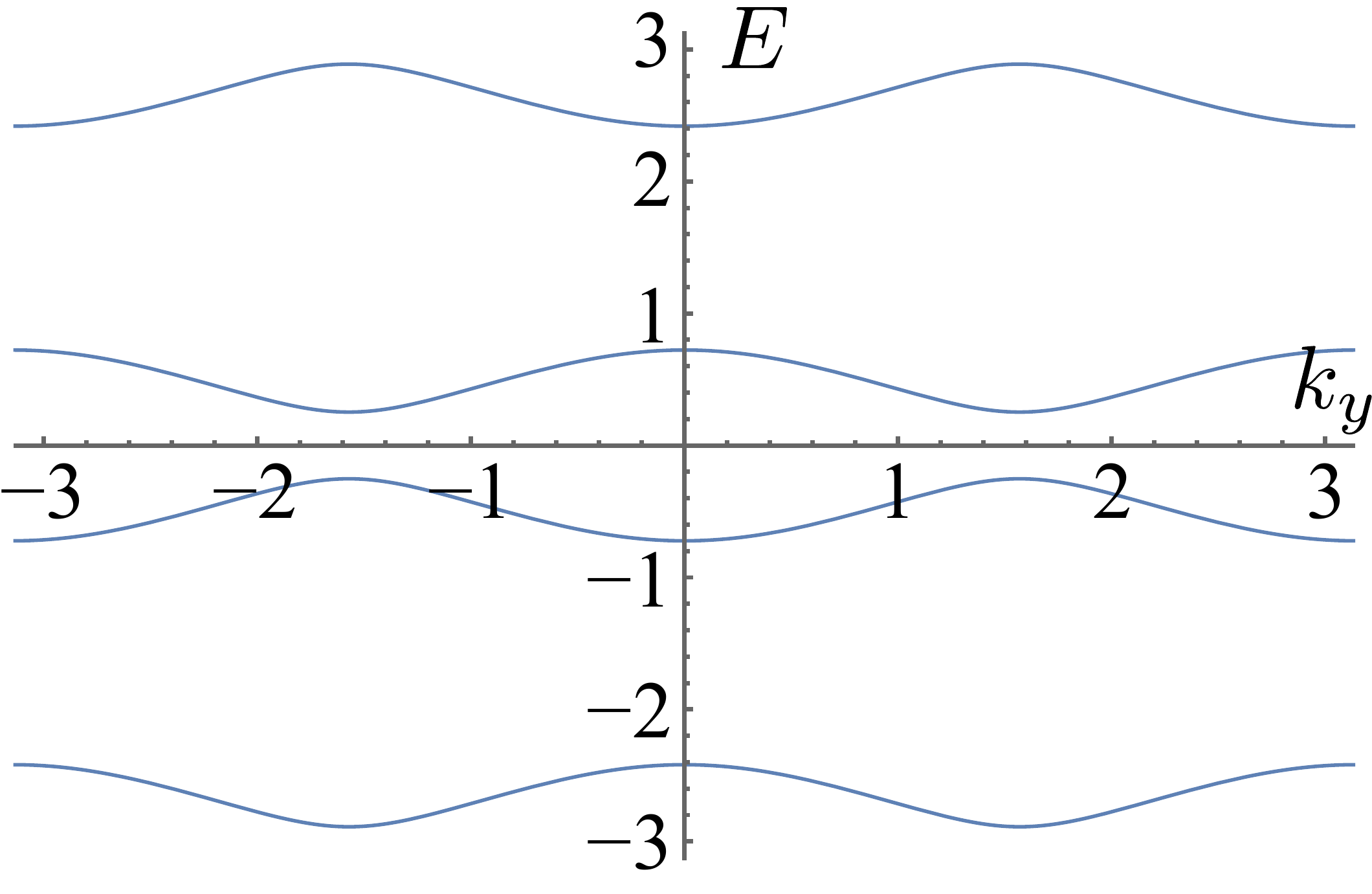}
\\
(n)
\end{minipage}
\hfill
\begin{minipage}[t]{0.32\textwidth}
\centering
\includegraphics[width=\textwidth]{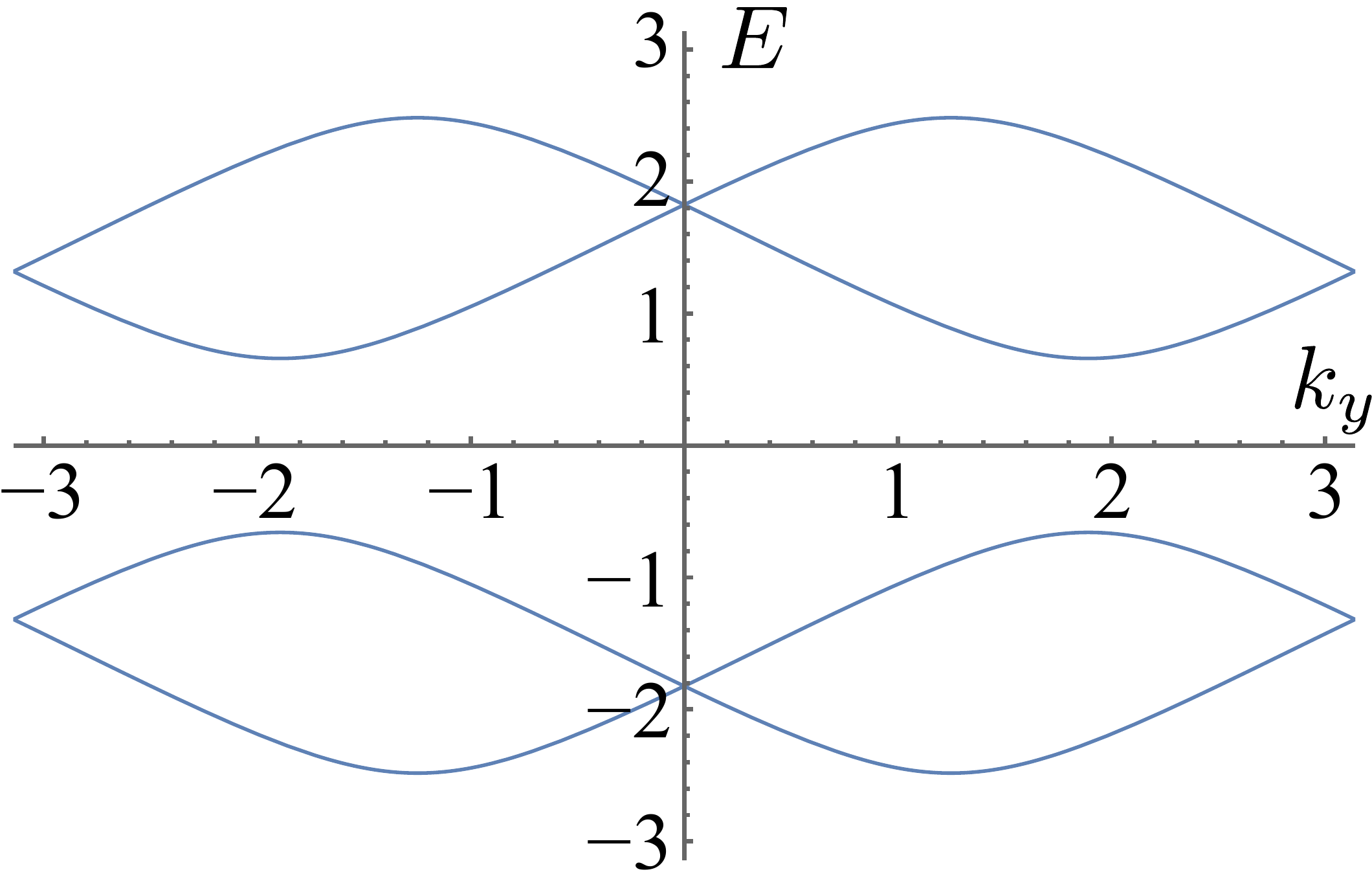}
\\
(o)
\end{minipage}
\caption{Cross sections of the bands at $k_x=0$, $k_x=\pi/2$ and $k_x=\pi$ on the first, second and third columns, respectively. We vary $\theta_y$ to $0$, $\pi/12$, $\pi/6$, $\pi/4$ and $\pi/3$ on the five rows from top to bottom, respectively. We set $\hbar=a=\Delta t=1$.}
\label{figS3}
\end{figure*}

Figure~\ref{figS3} shows the variation of the cross sections on $k_x=0$, $k_x=\pi/2$ and $k_x=\pi$ for $\theta_y=0,\pi/12,\pi/6,\pi/4,\pi/3$.
We can see the openings at each cross section become wider as we increase $\theta_y$.

\bibliographystyle{apsrev4-1}
\bibliography{yamagishi}

\end{document}